# Cloning in Elections: Finding the Possible Winners

**Edith Elkind**                                                    EELKIND@NTU.EDU.SG
*School of Physical and Mathematical Sciences*
*Nanyang Technological University, Singapore*

**Piotr Faliszewski**                                              FALISZEW@AGH.EDU.PL
*AGH University of Science and Technology*
*Krakow, Poland*

**Arkadii Slinko**                                          SLINKO@MATH.AUCKLAND.AC.NZ
*Department of Mathematics*
*University of Auckland, Auckland, New Zealand*

## Abstract

We consider the problem of manipulating elections by cloning candidates. In our model, a manipulator can replace each candidate $c$ by several *clones*, i.e., new candidates that are so similar to $c$ that each voter simply replaces $c$ in his vote with a block of these new candidates, ranked consecutively. The outcome of the resulting election may then depend on the number of clones as well as on how each voter orders the clones within the block. We formalize what it means for a cloning manipulation to be successful (which turns out to be a surprisingly delicate issue), and, for a number of common voting rules, characterize the preference profiles for which a successful cloning manipulation exists. We also consider the model where there is a cost associated with producing each clone, and study the complexity of finding a minimum-cost cloning manipulation. Finally, we compare cloning with two related problems: the problem of control by adding candidates and the problem of possible (co)winners when new alternatives can join.

## 1. Introduction

In many real-life elections, some of the candidates may have fairly similar positions on major issues, yet disagree on what is the best way to implement their common goals. Under many voting rules, and most glaringly under Plurality voting, these candidates run the risk of splitting the vote and losing to a candidate with an opposing program. This phenomenon can be exploited to alter the election outcome. For instance, New York Times wrote about a Republican political operative who recruited drifters and homeless people onto the Green Party ballot and freely admitted that these candidacies may siphon some support from the Democrats and therefore help Republicans (Lacey, 2010).

Such scenarios have been extensively studied in the (computational) social choice literature (see Section 6 for an overview). Depending on whether the manipulation is contemplated by one of the candidates or by an external party, this issue is known as *strategic candidacy problem* (this term was coined in Dutta, Jackson, & Le Breton, 2001, 2002) or the problem of *control by adding candidates*.

In this paper we will address a variant of this problem that is known as *cloning*. It is characterized by the following feature: each new candidate must be very similar to one of the existing candidates. This form of manipulative behavior was first identified and studied by Tideman (1987), who also gave a now classic example of a cloning strategy. Tideman wrote: "When I was 12 years old I was nominated to be treasurer of my class at school. A girl named Michelle was also nominated. I relished the prospect of being treasurer, so I made a quick calculation and nominated Michelle's best friend, Charlotte. In the ensuing election I received 13 votes, Michelle received 12, and Charlotte received 11, so I became treasurer" (Tideman, 1987, p. 1). The calculation was that, being friends, Michelle and Charlotte are 'similar' and that their electorate will be split.

In Tideman's example, the cloned alternative lost the election. However, one can also imagine scenarios where by cloning an alternative we increase its chances of winning. For example, suppose that an electronics website runs a competition for the best digital camera by asking consumers to vote for their two favorite models from a given list. If the list contains one model of each brand, and 60% of the consumers prefer Sony to Nikon to Kodak, while the remaining consumers prefer Kodak to Nikon to Sony, then Nikon will win the competition. On the other hand, if Sony is represented by two similar models, then the "Sony" customers are likely to vote for these two models of Sony, and the competition will be won by a Sony camera.

Just as in the general candidate addition scenario, cloning presents an opportunity for a party that is interested in manipulating the outcome of a preference aggregation procedure, such as an election or a consumer survey. Such a party—most likely, a campaign manager for one of the candidates—may invest in creating "clones" of one or more alternatives in order to make its most preferred alternative (or one of its "clones") win the election. This campaign management strategy has certain advantages over introducing an entirely new candidate: In the latter case, it may be hard to predict how the voters rank the new candidate, so the campaign manager would have to either invest in eliciting the new candidate's rankings, or be prepared to deal with rankings that differ from her initial expectations. In comparison, the outcomes of cloning are much more predictable, and therefore manipulation by cloning may be easier to implement. A natural question, then, is which voting rules are resistant to such manipulation, and whether the manipulator can compute the optimal cloning strategy for a given election in a reasonable amount of time.

As mentioned above, the first study of cloning was undertaken by Tideman (1987), who introduced the concept of independence of clones as a criterion for voting rules. Apparently unaware of Tideman's work, Laffond, Laine, and Laslier (1996) introduced the notion of *composition consistency*, which is an analogue of independence of clones for tournament solutions (Laslier, 1997), i.e., voting rules that are defined on the majority relation that corresponds to the voters' preferences. Later, Laslier (1996, 2000) introduced the notion of *cloning consistency*, which is equivalent to independence of clones. We will discuss these results in more detail in Section 6.

In all these papers, the authors concentrated on finding out whether a certain voting rule is independent of clones or on constructing new rules with this property. In our work, we take a somewhat different perspective: Instead of looking at cloning as a manipulative action that should be prevented, we view cloning as a campaign management tool. This point of view raises a number of questions that have not been considered—or have been considered from a different angle—in the previous work:

**What does it mean for cloning to be successful?** We assume that the campaign manager can produce clones of existing candidates, and the voters rank them in response.

We assume that clones are similar enough to be ranked as a group by each voter; however, the order of clones in such groups may differ from one voter to another. Since the campaign manager cannot control or predict the order of clones in each voter's ranking, we assume that this order is random, i.e., each voter ranks the cloned candidates in each possible order with the same probability; indeed, this would be the case if all clones were indistinguishable. In this probabilistic model, each cloning strategy succeeds with a certain probability. Let $q$ be some real number between 0 and 1. We say that manipulation by cloning is *q-successful* if the probability of electing a desired candidate $p$ is at least $q$. We focus on two extreme cases: (1) $p$ wins under each possible ordering of the clones, and (2) $p$ wins under at least one ordering of the clones. In case (1), the cloning is 1-successful; in case (2), following the notation typically used when dealing with limits in continuous mathematics, we will say that the cloning is $0^+$-successful.

**In which instances of elections can cloning be successful?** While the previous work shows that many well-known voting rules are susceptible to cloning, no attempt has been made to characterize the elections in which a specific candidate can be made a winner with respect to a given voting rule by means of cloning. However, from the point of view of a campaign manager who considers cloning as one of the ways to run the campaign, it would be important to know if he can change the outcome of a given election by a cloning manipulation. Thus, in this paper we provide such characterization results for several prominent voting rules. Often, the candidates for which a successful cloning manipulation exists can be characterized in terms of well-known notions of social choice such as Pareto undominated alternative (a candidate $c$ is Pareto undominated if for every other candidate $c'$ there is a voter that prefers $c$ to $c'$), Condorcet loser (candidate $c$ is a Condorcet loser if for every other candidate $c'$ more than half of the voters prefer $c'$ and $c$), or Uncovered Set (see Section 5.3 or Miller, 1977; Fishburn, 1977; Laslier, 1997).

**Which candidates can be cloned and to what extent?** The existing research on cloning does not place any restrictions on the number of clones that can be introduced, or on the identities of the candidates that can be cloned. On the other hand, it is clear that in practical campaign management scenarios these issues cannot be ignored; not all candidates can be cloned, and creating a clone of a given candidate may be costly. Thus, we consider settings in which each clone of each candidate comes at a cost, and we seek a least expensive successful cloning strategy. We will mostly focus on the standard model where clones come at zero cost, and on the *unit cost* model, where each clone has the same cost.

**What is the computational complexity of finding cloning strategies?** Finally, we investigate the computational complexity of finding successful cloning strategies. In practice, it is not sufficient to know that cloning *might* work: We need to know exactly which strategy to use. We believe that our paper is the first to consider the computational aspect of cloning. Following the line of work initiated by the seminal papers of Bartholdi, Tovey, and Trick (1989, 1992), we seek to classify prominent voting rules according to whether they admit efficient algorithms for finding a cloning manipulation.

One might argue that in real-life elections cloning is not a practical campaign management tool: After all, recruiting a new candidate that is sufficiently similar to the existing ones may be very difficult, if not impossible. Nonetheless, there are natural scenarios where our model of cloning is practical and well-motivated. Below, we provide two such examples.

First, let us consider an election in which parties nominate candidates for some position, and each party can nominate several candidates. The voters, especially those not following the political scene closely, are likely to perceive candidates who belong to the same party as clones. A party's campaign manager might attempt to strategically choose the number of candidates nominated by her party. In fact, she might even be able to affect the number of candidates nominated by other parties (e.g., by accusing them of not giving the voters enough choice).

Second, let us consider an environment where, as suggested by Ephrati and Rosenschein (1997), software agents vote to choose a joint plan (that is, the alternatives are possible joint plans or steps of possible joint plans). In such a system, the agents can easily come up with minor variations of the (steps of the) plan, effectively creating clones of the candidates. (A very similar example regarding a society of agents choosing a project to implement is given in Laslier, 1996).

In both cases it is reasonable to assume that all clones are ranked contiguously and the cost of creating each clone is the same; moreover, to be successful, a cloning strategy should be easy to compute. Therefore our model provides a good fit for these scenarios.

## 2. Preliminaries

We start by presenting a brief overview of the social choice concepts that will be used in this paper; we point the reader to the book of Arrow, Sen, and Suzumura (2002) for additional background.

Given a set $A$ of *alternatives* (also called *candidates*), a voter's *preference* $R$ is a *linear order* over $A$, i.e., a total transitive antisymmetric binary relation over $A$. An *election* $E$ with $n$ voters is given by its set of alternatives $A$ and a *preference profile* $\mathcal{R} = (R_1, \ldots, R_n)$, where $R_i$ is the preference of voter $i$; we write $E = (A, \mathcal{R})$. For readability, we sometimes write $\succ_i$ in place of $R_i$. Sometimes when specifying a preference order $R_i$ we write $X \succ_i Y$, where $X$ and $Y$ are two disjoint subsets of $A$. This notation means that each member of $X$ is preferred to each member of $Y$ but the relative ordering of candidates within $X$ and within $Y$ is irrelevant to the discussion (unless specified separately). Also, we denote by $|\mathcal{R}|$ the number of voters in the election. Given an election $E = (A, \mathcal{R})$, we say that an alternative $c \in A$ is *Pareto undominated* if for each alternative $c' \in A$, $c' \neq c$, at least one voter ranks $c$ ahead of $c'$. Given two candidates $a, c \in A$, we set $W(c, a) = |\{i \mid c \succ_i a\}|$. We say that $c$ *beats* $a$ in their *pairwise contest* if $W(c, a) > W(a, c)$; if $W(c, a) = W(a, c)$, the pairwise contest between $a$ and $c$ is said to be *tied*.

A voting rule $\mathcal{F}$ is often defined as a mapping from elections with a fixed set of alternatives $A$ to the set $2^A \setminus \{\emptyset\}$ of all nonempty subsets of $A$. However, in this work, we are interested in situations where the number of alternatives may change. Thus, we require voting rules to be defined for arbitrary finite sets of alternatives and preference profiles over those alternatives. We say that a *voting rule* $\mathcal{F}$ is a mapping from pairs of the form $E = (A, \mathcal{R})$, where $A$ is some finite set and $\mathcal{R}$ is a preference profile over $A$, to nonempty subsets of $A$. The elements of $\mathcal{F}(E)$ are called the *winners* of the election $E$. Thus, we allow an election to have more than one winner, i.e., we work with social choice correspondences; this model is also called the *non-unique winner* model.

In this paper we consider the following voting rules (for all rules described in terms of scores the winners are the alternatives with the maximum score):

**Plurality.** The *Plurality score* $Sc_P(c)$ of a candidate $c \in A$ is the number of voters that rank $c$ first.

**Veto.** The *Veto score* $Sc_V(c)$ of a candidate $c \in A$ is the number of voters that do not rank $c$ last.

**Borda.** Given an election $(A, \mathcal{R})$ with $|\mathcal{R}| = n$, the *Borda score* $Sc_B(c)$ of a candidate $c \in A$ is given by $Sc_B(c) = \sum_{i=1}^{n} |\{a \in A \mid c \succ_i a\}|$.

**$k$-Approval.** For any $k \geq 1$, the *$k$-Approval score* $Sc_k(c)$ of a candidate $c \in A$ is the number of voters that rank $c$ in the top $k$ positions. Plurality is simply 1-Approval.

**Plurality with Runoff.** In the first stage, all but two candidates with the top two Plurality scores are eliminated. Then, we run a pairwise contest between the two survivors; the winner(s) are the candidate(s) who get at least $\lceil |\mathcal{R}|/2 \rceil$ votes at this stage, i.e., win or tie the pairwise contest. We may need to break a tie after the first stage if there are more than two candidates with the maximum Plurality score, or one top-scorer and several candidates with the second-best Plurality score. To this end we use the *parallel universes* tie-breaking rule (Conitzer, Rognlie, & Xia, 2009): a candidate $c$ is considered to be a winner if he wins or ties in the runoff for *some* way of breaking ties after the first stage.

**Maximin.** The *Maximin score* $Sc_M(c)$ of a candidate $c \in A$ is $Sc_M(c) = \min_{a \in A} W(c, a)$, i.e., it is the number of votes that $c$ gets in his worst pairwise contest.

**Copeland.** The *Copeland score* $Sc_C(c)$ of a candidate $c \in A$ is $|\{a \mid W(c, a) > W(a, c)\}| - |\{a \mid W(a, c) > W(c, a)\}|$. That is, $c$ receives 1 point for each pairwise contest she wins, 0 points for a tie, and $-1$ point for each pairwise contest she loses. This is equivalent to the more conventional definition, in which, for each candidate $a$, $c$ gets 1 point if she wins the pairwise contest against $a$, 0.5 points if there is a tie, and 0 if she loses the contest.

A candidate $c$ is a *Condorcet winner* (respectively, *Condorcet loser*) if for each other candidate $a$ it holds that $W(c, a) > W(a, c)$ (respectively, $W(c, a) < W(a, c)$). Naturally, not every election has a Condorcet winner or a Condorcet loser. Observe that in a Copeland election with $m$ candidates the score of a Condorcet winner is $m - 1$ and the score of a Condorcet loser is $-(m-1)$.

Many results of this paper are computational and thus we assume that the reader is familiar with the standard notions of computational complexity such as classes P and NP, many-one reductions, NP-hardness and NP-completeness. Most of our NP-hardness results follow by reductions from the Exact Cover by 3-Sets problem, defined below.

**Definition 2.1** (Garey & Johnson, 1979). *An instance $(G, \mathcal{S})$ of* EXACT COVER BY 3-SETS (X3C) *is given by a ground set $G = \{g_1, \ldots, g_{3K}\}$ and a family $\mathcal{S} = \{S_1, \ldots, S_M\}$ of subsets of $G$, where $|S_i| = 3$ for each $i = 1, \ldots, M$. It is a "yes"-instance if there is a subfamily $\mathcal{S}' \subseteq \mathcal{S}$, $|\mathcal{S}'| = K$, such that for each $g_i \in G$ there is an $S_j \in \mathcal{S}'$ such that $g_i \in S_j$, and a "no"-instance otherwise.*

## 3. Our Framework

Cloning and independence of clones were previously defined by Laslier (2000), Tideman (1987), and Zavist and Tideman (1989). However, we need to modify the definition given in these papers in order to model the manipulator's intentions and the budget constraints. We will now describe our model formally.

**Definition 3.1.** *Let $E = (A, (R_1, \ldots, R_n))$ be an election with a set of candidates $A = \{c_1, \ldots, c_m\}$. We say that an election $E' = (A', (R_1', \ldots, R_n'))$ is obtained from $E$ by replacing a candidate $c_j \in A$ with $k$ clones for some $k > 0$ if $A' = (A \setminus \{c_j\}) \cup \{c_j^{(1)}, \ldots, c_j^{(k)}\}$ and for each $i \in [n]$, $R_i'$ is a total order over $A'$ such that for any $a \in A \setminus \{c_j\}$ and any $s \in [k]$ it holds that $c_j^{(s)} \succ_i' a$ if and only if $c_j \succ_i a$.*

*We say that an election $E^* = (A^*, \mathcal{R}^*)$ is cloned from an election $E = (A, \mathcal{R})$ if there is a vector of nonnegative integers $(k_1, \ldots, k_m)$ such that $E^*$ is derived from $E$ by replacing each $c_j$, $j = 1, \ldots, m$, with $k_j$ clones.*

Thus, when we clone a candidate $c$, we replace her with a group of new candidates that are ranked together in all voters' preferences. Observe that according to the definition above, cloning a candidate $c_j$ once means simply changing his name to $c_j^{(1)}$ rather than producing an additional copy of $c_j$. While not completely intuitive, this choice of terminology simplifies some of the arguments in the rest of the paper.

The definition above is essentially equivalent to the one given by Zavist and Tideman (1989); the main difference is that we explicitly model cloning of more than one candidate. However, we still need to introduce the two other components of our model: a definition of what it means for a cloning to be successful, and the budget.

We start with the former, assuming throughout this discussion that the voting rule is fixed. Observe that the final outcome of cloning depends on the relative ranking of the clones chosen by each voter, which is typically not under the manipulator's control.[1] Thus, a cloning may succeed for some orderings of the clones, but not for others. The election authorities may approach this issue from the worst-case perspective, and consider it unacceptable when a given cloning succeeds for *at least one* ordering of clones by the voters. Alternatively, they can take an average-case perspective, i.e., assume that the voters rank the clones randomly and independently, with each ordering of the clones being equally likely (due to similarities among the clones), and consider it acceptable for a cloning manipulation to succeed with probability that does not exceed a certain threshold. Similarly, while an extremely cautious manipulator would view cloning as successful only if it succeeds for *all* orderings, a more practically-minded one would be happy with a cloning that succeeds with high probability. We will now present a definition that captures all of these attitudes.

---

1. Our general model, and, specifically, the notion of $0^+$-successful cloning (to be defined in a few paragraphs), captures the situation where the manipulator does have full control over the ordering of the clones in voters' preferences.

**Definition 3.2.** *Given a positive real $q$, $0 < q \leq 1$, we say that the manipulation by cloning (or simply cloning) is $q$-successful if (a) the manipulator's preferred candidate is not a winner of the original election, and (b) a clone of the manipulator's preferred candidate is a winner of the cloned election with probability at least $q$.*

The two worst-case approaches discussed above are special cases of this framework. Indeed, a cloning succeeds for all orderings if and only if it is 1-successful. Similarly, it succeeds for some ordering if and only if it is $q$-successful for some $q > 0$ (where $q$ may depend on $|A|$ and $|\mathcal{R}|$). In the latter case, we will say that the cloning is $0^+$-*successful*; this is equivalent to saying that the manipulator would succeed if he could dictate each voter how to order the clones. We will use this equivalent formulation very often as it simplifies proofs.

Observe that, according to our definition, the manipulator succeeds as long as *any one* of the clones of the preferred candidate wins. This assumption is natural if the clones represent the same company (e.g., Coke Light and Coke Zero) or political party. However, if a campaign manager has created a clone of his candidate simply by recruiting an independent candidate to run on a similar platform, he may find the outcome in which this new candidate wins less than optimal. We could instead define success as a victory by the original candidate $c$ (i.e., the clone $c^{(1)}$), but, at least for neutral voting rules, this is equivalent to scaling down the success threshold $q$ by a factor of $k$, where $k$ is the number of clones of the preferred candidate. Indeed, any preference profile in which the original candidate wins can be transformed into one in which some clone wins, by switching their order in each voter's preferences, so $c^{(1)}$ wins with the same probability as any other clone. In particular, this means that under this definition a cloning manipulation can only be 1-successful if it does not clone the manipulator's preferred candidate.

Another issue that we need to address is that of the costs associated with cloning. Indeed, the costs are an important aspect of realistic campaign management, as the manager is always restricted by the budget of the campaign. The most general way to model the cloning costs for an election with the initial set of candidates $A = \{c_1, \ldots, c_m\}$ is by introducing a *cost function* $p \colon [m] \times \mathbb{Z}^+ \to \mathbb{Z}^+ \cup \{0, +\infty\}$, where $p(i, j)$ denotes the cost of producing the $j$-th copy of candidate $c_i$. Note that $p(i, 1)$ corresponds to not producing additional copies of $i$, so we require $p(i, 1) = 0$ for all $i \in [m]$. We remark that it is natural to assume that all costs are nonnegative (though some of them may equal zero), whereas the assumption that all costs are integer-valued is made for computational reasons; this is not a real restriction as monetary values are discrete.

We assume that the marginal cost of introducing an additional cloned candidate eventually becomes constant, that is, there exists a $t > 1$ such that $p(i, j) = p(i, t)$ for each $j > t$. This ensures that the cost function has a finite representation; specifically, we can encode $p$ as an $m$-by-$t$ table with entries in $\mathbb{Z}^+ \cup \{0, +\infty\}$.

**Definition 3.3.** *An instance of the $q$-CLONING problem for $q \in \{0^+\} \cup (0, 1]$ is given by the initial set of candidates $A = \{c_1, \ldots, c_m\}$, a preference profile $\mathcal{R}$ over $A$ with $|\mathcal{R}| = n$, a manipulator's preferred candidate $c \in A$, a parameter $t > 1$, a cost function $p \colon [m] \times [t] \to$* $\mathbb{Z}^+ \cup \{0, +\infty\}$ *(with the interpretation that $p(i, j) = p(i, t)$ for all $i = 1, \ldots, m$ and all $j > t$), a budget $B$, and a voting rule $\mathcal{F}$. We ask if there exists a $q$-successful cloning with respect to $\mathcal{F}$ that costs at most $B$.*

For many of the voting rules that we consider, it is easy to bound the number of clones needed for $0^+$-successful or $1$-successful cloning (if one exists), and this bound is usually polynomial in $n$ and $m$. Thus, $t$ can often be assumed to be polynomial in $n$ and $m$.

We focus on two natural special cases of $q$-CLONING:

1. ZERO COST (ZC): $p(i, j) = 0$ for all $i \in [m]$, $j \in \mathbb{Z}^+$. In this case we would like to decide if an election is manipulable by cloning when money is not a concern.

2. UNIT COST (UC): $p(i, j) = 1$ for all $i \in [m]$, $j \geq 2$. This model assumes that creating each new clone has a fixed cost that is equal for all candidates.

We say that an election $E$ is *$q$-manipulable by cloning with respect to a voting rule $\mathcal{F}$* if it admits a $q$-successful manipulation by cloning with respect to $\mathcal{F}$ in the ZC model.

In the rest of the paper, we characterize the $q$-manipulable elections and discuss the complexity of the $q$-CLONING problem for a number of well-known voting rules, focusing on the ZC and UC models. Clearly, hardness results for these special cases also imply hardness results for the general cost model. Similarly, hardness results for the ZC $q$-CLONING imply hardness results for UC $q$-CLONING; to reduce ZC $q$-CLONING to UC $q$-CLONING it suffices to set $B = +\infty$. We emphasize that whenever we say that $q$-CLONING is easy, we refer to the general cost model; in contrast, $q$-manipulability refers to susceptibility to cloning manipulation with zero costs and/or unlimited budget.

## 4. Prominent Voting Rules for Which Cloning is Easy

In this section we study $q$-CLONING for Plurality, Plurality with Runoff, Veto, and Maximin. Surprisingly, these four rules exhibit very similar behavior with respect to cloning.

### 4.1 Plurality

We start by considering Plurality, which is arguably the simplest voting rule.

**Theorem 4.1.** *An election is $0^+$-manipulable by cloning with respect to Plurality if and only if the manipulator's preferred candidate $c$ does not win, but is ranked first by at least one voter. Moreover, for Plurality $0^+$-CLONING can be solved in linear time.*

*Proof.* Clearly, if $c$ has no first-place votes, no cloning manipulation can make her a winner. Now, assume that $c$'s Plurality score $Sc_P(c)$ is at least 1, and let $C = \{a \in A \mid Sc_P(a) > Sc_P(c)\}$ denote the set of candidates whose Plurality score is greater than that of $c$. For each $a \in C$, we create $k_a = \lceil \frac{Sc_P(a)}{Sc_P(c)} \rceil$ clones of $a$.

Recall that to show that a cloning is $0^+$-successful, we only need to specify an ordering of the clones that makes $c$ a winner. One such ordering can be obtained as follows. For each $a \in C$, let $\mathcal{R}^a$ denote the set of orders in which $a$ is ranked first. Split $\mathcal{R}^a$ into $k_a$ groups, where the first $k_a - 1$ groups have size $Sc_P(c)$, while the last group has size $Sc_P(a) - (k_a - 1)Sc_P(c) \leq Sc_P(c)$. Let the voters in the $i$-th group rank the $i$-th clone of $a$ first, followed by the rest of the clones in an arbitrary order. Under this ordering of the clones, the Plurality score of each candidate is at most $Sc_P(c)$, so $c$ is a winner, i.e., our cloning is $0^+$-successful.

To prove the second statement of the theorem, note that the above-described algorithm for finding a $0^+$-successful cloning is optimal: we can reduce a candidate's Plurality score only by cloning this candidate, and if a candidate $a$ with score $Sc_P(a) > Sc_P(c)$ is cloned less than $k_a$ times, at least one of its clones will obtain more than $Sc_P(c)$ Plurality votes. Thus, we simply need to compute the cost of cloning each candidate $a \in C$ exactly $k_a$ times, and compare it with our budget $B$. □

It is not too hard to strengthen the first statement of Theorem 4.1 from $0^+$-manipulability to $q$-manipulability for any $q \in (0, 1)$.

**Theorem 4.2.** *For any $q \in (0, 1)$, a Plurality election is $q$-manipulable by cloning if and only if the manipulator's preferred candidate $c$ does not win, but is ranked first by at least one voter. However, no Plurality election is 1-manipulable by cloning.*

*Proof.* Fix $q \in (0, 1)$. Suppose that the manipulator's preferred candidate $c$ does not win, but $Sc_P(c) > 0$. For each candidate $a \in A$ with $Sc_P(a) > Sc_P(c)$, we set $s = Sc_P(a)$ and create $k = \binom{s}{2} \lceil \frac{m-1}{1-q} \rceil$ clones of $a$. Then the probability that one of the clones of $a$ will be top-ranked two or more times is at most $k \cdot \binom{s}{2} \cdot \frac{1}{k^2} \leq \frac{1-q}{m-1}$. By the union bound, with probability at least $q$ none of the newly introduced clones gets more than one vote. On the other hand, none of the uncloned candidates had more Plurality votes than $c$, and cloning did not change that. Therefore, $c$ is among the winners of the resulting election. However, obviously the manipulator cannot make $c$ a winner with probability 1: if all voters order the clones in the same way, the "most popular" clone of each candidate will have the same Plurality score as the original candidate. □

The procedure described in the proof of Theorem 4.2 introduces at most $(m-1)\binom{n}{2} \lceil \frac{m-1}{1-q} \rceil$ clones; this number is polynomial in $m$ and $n$ for any constant $q$. However, this number of clones is not necessarily optimal, i.e., our procedure is *not* a polynomial-time algorithm for $q$-Cloning. In fact, the complexity of Plurality $q$-Cloning for $q \in (0, 1)$ remains an open problem.

## 4.2 Veto and Plurality with Runoff

The Veto rule exhibits extreme vulnerability to cloning.

**Theorem 4.3.** *Any election in which the manipulator's preferred candidate $c$ does not win is 1-manipulable by cloning with respect to Veto. Moreover, for Veto both $0^+$-Cloning and 1-Cloning can be solved in linear time.*

*Proof.* Consider a profile in which the Veto score of the manipulator's preferred candidate $c$ is $k$. This means that $c$ is ranked last $n - k$ times. If we clone $c$ at least $n - k + 1$ times, we are guaranteed that at least one clone of $c$ is never ranked last, so it is among the Veto winners.

For $0^+$-CLONING, observe that it is not useful to clone candidates other than $c$. Thus, the optimal solution is to make two copies of $c$ and ask all voters to order them in the same way. Then the "better" clone is never ranked last and is among the winners.

For 1-CLONING, let $\ell$ be the Veto score of the election winner(s), and let $k$ be the Veto score of $c$. As argued above, we do not benefit from cloning candidates other than $c$, so we only need to determine the optimal number of $c$'s clones. Suppose first that $\ell = n$. Then, by the argument above, it is sufficient to create $n - k + 1$ clones of $c$. It is easy to see that this number is also necessary: if there are $n - k$ or fewer clones, each of them may be ranked last by some voter.

Now, suppose that $\ell < n$. Let $\ell' = n - \ell$, $k' = n - k$, set $r = \lfloor \frac{k'}{\ell'+1} \rfloor$, and create $r + 1$ clones of $c$. This number of clones is clearly sufficient: we have $r + 1 > \frac{k'}{\ell'+1}$, so if each clone of $c$ is ranked last at least $\ell' + 1$ times, then the total number of voters that rank $c$ last in the original profile would be at least $(\ell' + 1)(r + 1) > k'$, a contradiction. On the other hand, it is necessary to introduce at least $r + 1$ clones. Indeed, if there are only $r$ clones, we can split the voters that rank $c$ last into $r$ groups, where the first $r - 1$ groups have size $\ell' + 1$, the last group has size $k' - (r - 1)(\ell' + 1) \geq \ell' + 1$ (where the inequality follows from $r \leq \frac{k'}{\ell'+1}$), and the voters in the $i$-th group rank the $i$-th clone of $c$ last. In this preference profile, each clone of $c$ is vetoed as least $\ell' + 1$ times and therefore is not among the winners. $\qquad\square$

We now consider Plurality with Runoff.

**Theorem 4.4.** *An election is $0^+$-manipulable by cloning with respect to Plurality with Runoff if and only if the manipulator's preferred candidate $c$ is not a current winner, and either*

*(1) $Sc_P(c) \geq 2$, or*

*(2) $Sc_P(c) = 1$ and $c$ wins or ties its pairwise contest against some alternative $w$ whose Plurality score is strictly positive.*

*Moreover, for Plurality with Runoff $0^+$-CLONING can be solved in polynomial time.*

*Proof.* Suppose first that $Sc_P(c) \geq 2$. We can introduce two clones of $c$, which we will denote by $c^{(1)}$ and $c^{(2)}$, split the voters who rank $c$ first into two nonempty groups, ask the voters in the first group to rank $c^{(1)}$ first, and ask the voters in the second group to rank $c^{(2)}$ first. Next, for each candidate $a \neq c$ we create $Sc_P(a)$ clones of $a$, and ask the $i$-th voter among those who rank $a$ first to rank the $i$-th clone of $a$ first. In the resulting election, the Plurality score of both $c^{(1)}$ and $c^{(2)}$ is at least 1, and the Plurality score of any other candidate is at most 1. Thus, there is a way to break ties after the first round so that both $c^{(1)}$ and $c^{(2)}$ progress to the final round, and one of them wins. As we apply the parallel universes tie-breaking rule, this means that this cloning is $0^+$-successful.

If $Sc_P(c) = 1$, we use the same strategy as in the previous case, except that we do not clone $c$. In the resulting election the Plurality score of any candidate is at most 1. Let $w$ be an alternative with nonzero Plurality score that loses or ties its pairwise election against $c$. It is easy to see that any clone of $w$ also loses or ties its pairwise election against $c$. Since $w$'s Plurality score in the original election is positive, there exists a parallel universe where $c$ and a clone of $w$ meet in the final, which means that $c$ is a winner of the resulting election.

Thus, we see that the conditions of the theorem are sufficient. We will now show that they are necessary. First, note that if $c$ is not ranked first by any voter, then $c$ will not win irrespective of which candidates we clone. Indeed, if in the cloned election there are two candidates with nonzero Plurality scores, $c$ will not reach the final, and if all voters rank some candidate $a \neq c$ first, $c$ may reach the final, but will lose to $a$ in the final. Now, suppose that $Sc_P(c) = 1$, but any alternative $w$ with $Sc_P(w) > 0$ beats $c$ in their pairwise contest; as argued above this also holds for any clone of $w$. Since $c$ is not a winner prior to cloning, there exists at least one other candidate with nonzero Plurality score, and therefore at most one clone of $c$ can progress to the second round; hence, there is no benefit to cloning $c$. Moreover, even if $c$ reaches the second round, he has to face (a clone of) some alternative with nonzero Plurality score in the original election, and by our assumption any such clone would beat $c$ in the final.

To complete the proof of the theorem, it remains to give a polynomial-time algorithm for $0^+$-CLONING under Plurality with Runoff. As suggested by the discussion above, there are two ways to make $c$ a winner: we can either (1) try to clone $c$ (and possibly some other alternatives) in order to ensure that only clones of $c$ go to the runoff, or (2) try to clone alternatives other than $c$ only, to ensure that $c$ goes to the runoff against an alternative that she can defeat. Our algorithm implements both options and accepts if either of them is within the budget.

For option (1), we introduce two clones of $c$, which we will denote by $c^{(1)}$ and $c^{(2)}$ (clearly, there is no benefit to creating more than two clones of $c$). It is not hard to see that our optimal strategy is to ask the voters to order these clones so that they get (almost) identical Plurality scores, i.e., so that the Plurality scores of $c^{(1)}$ and $c^{(2)}$ are $\lceil \frac{Sc_P(c)}{2} \rceil$ and $\lfloor \frac{Sc_P(c)}{2} \rfloor$, respectively. Let $k = \lfloor \frac{Sc_P(c)}{2} \rfloor$, and let $C = \{a \in A \setminus \{c\} \mid Sc_P(a) > k\}$ denote the set of candidates whose Plurality score is greater than that of $c^{(2)}$. For each candidate $a \in C$, we introduce $k_a = \lceil \frac{Sc_P(a)}{k} \rceil$ clones of $a$. With nonzero probability, this action ensures that the Plurality score of each alternative except $c^{(1)}$ and $c^{(2)}$ does not exceed $k$, and thus there is a parallel universe in which $c^{(1)}$ competes against $c^{(2)}$ in the runoff and one them wins. It is easy to see that this strategy gives the cheapest way of implementing option (1).

Option (2) can be used only if candidate $c$ wins or ties her pairwise contest against at least one other candidate with nonzero Plurality score. Let $D = \{a \in A \setminus \{c\} \mid Sc_P(a) \geq 1$ and $W(c, a) > W(a, c)\}$. For each $w \in D$ we compute the cost of the manipulation that results in $c$ competing against $w$ in the runoff in one of the parallel universes; we then pick the cheapest of these manipulations. It remains to explain how to compute such a manipulation for a specific $w \in D$.

Let $k = \min\{Sc_P(c), Sc_P(w)\}$, and let $C = \{a \in A \setminus \{c, w\} \mid Sc_P(a) > k\}$. If not cloned, the candidates in $C$ can prevent $c$ and $w$ from meeting each other in the final round. Thus, we create $k_a = \lceil \frac{Sc_P(a)}{k} \rceil$ clones of each $a \in C$. It is easy to see that this manipulation results in a nonzero chance of $c$ winning and that we cannot produce fewer clones for a given $w \in D$. This completes the proof. $\qquad \square$

We can also characterize elections that are $q$-manipulable with respect to Plurality with Runoff for $q \in (0, 1]$.

**Theorem 4.5.** *For any $q \in (0, 1)$, an election is $q$-manipulable by cloning with respect to Plurality with Runoff if and only if it is $0^+$-manipulable by cloning with respect to it. However, no election is 1-manipulable.*

*Proof.* It is immediate that under Plurality with Runoff no election is 1-manipulable: if our preferred candidate is not a winner and after cloning all voters rank the clones identically, then our preferred candidate still loses.

To see that for each $q \in (0, 1)$ an election is $q$-manipulable with respect to Plurality with Runoff if and only if it is $0^+$-manipulable with respect to it, it suffices to combine the proofs of Theorems 4.4 and 4.2. In more detail, the proof of Theorem 4.2 explains how to clone some of the candidates so that with probability at least $q$ the Plurality scores of all candidates in the resulting profile $\mathcal{R}'$ do not exceed 1. We will now argue that whenever this happens (i.e., with probability $q$) $c$ is a Plurality with Runoff winner as long as he satisfies the conditions of Theorem 4.4.

Indeed, suppose that in $\mathcal{R}'$ the Plurality score of each candidate is at most 1. Then if $c$'s Plurality score in the original election is at least 2, it has to be the case that at least two clones of $c$ have positive Plurality scores in $\mathcal{R}'$, and hence there is a parallel universe where they meet in the final. On the other hand, if $c$'s Plurality score in the original election is 1, but he beats or ties some candidate $w$ with nonzero Plurality score in a pairwise election, for $\mathcal{R}'$ there is a parallel universe where $c$ meets (a clone of) $w$ in the final. Thus, $c$ is a Plurality with Runoff winner in $\mathcal{R}'$, and the proof is complete. $\square$

We remark that the cloning strategy presented in the proof of Proposition 4.5 is, in a sense, degenerate: it operates in the same way irrespective of the identity of the preferred candidate $c$, and has the effect of making all "eligible" candidates the Plurality with Runoff winners. (This can be seen as an artifact of the non-unique winner model, where this outcome is viewed as acceptable. If we are interested in making $c$ the *unique* Plurality with Runoff winner, we may need a more sophisticated cloning strategy; however, this question is outside of the scope of this paper.) Observe also that under Plurality with Runoff cloning can be used to manipulate in favor of a Condorcet loser.[2]

**Example 4.6.** Let us consider the following Plurality with Runoff election: $A = \{a, b, c, d\}$, and there are 17 voters whose preference orders are:

$$c \succ a \succ b \succ d \qquad \text{8 voters}$$
$$a \succ b \succ d \succ c \qquad \text{3 voters}$$
$$b \succ a \succ d \succ c \qquad \text{3 voters}$$
$$d \succ a \succ b \succ c \qquad \text{3 voters}$$

Clearly, $c$ is a Condorcet loser. Further, if we apply Plurality with Runoff, $c$ gets into the second round, but then loses there. Yet if we produce two clones of $c$, it is possible that each of them receives four Plurality points, enters the runoff and, as a result, one of them wins the election.

––––––––––––––––––––
2. We are grateful to one of the JAIR referees for this example.

### 4.3 Maximin

Surprisingly, Maximin behaves in essentially the same way as Plurality with respect to cloning, i.e., by cloning a candidate, we can reduce its Maximin score to 1 with nonzero probability.

Consider the following election, which will be used in our constructions throughout this section. Let $E = (A, \mathcal{R})$ with $A = \{a_1, \ldots, a_k\}$, $\mathcal{R} = (R_1, \ldots, R_k)$, where for $i \in [k]$ the preferences of the $i$-th voter are given by $a_i \succ_i a_{i+1} \succ_i \ldots \succ_i a_k \succ_i a_1 \succ_i \ldots \succ_i a_{i-1}$. We will refer to any election that can be obtained from $E$ by renaming the candidates as a *$k$-cyclic election*. In this election, assuming $a_0 = a_k$, for each $i = 1, \ldots, k$ there are $k - 1$ voters that prefer $a_{i-1}$ to $a_i$ and 1 voter that prefers $a_i$ to $a_{i-1}$. Thus, the Maximin score of each candidate in $A$ is 1. Further, this remains true if we add arbitrary candidates to the election, no matter how the voters rank the additional candidates. This means that by cloning a candidate $a$ in an $n$-voter election $n$ times and telling the voters to order the $n$ clones of $a$ as in an $n$-cyclic election, we can reduce $a$'s Maximin score to 1.

**Theorem 4.7.** *An election is $0^+$-manipulable by cloning with respect to Maximin if and only if the manipulator's preferred candidate $c$ does not win, but is Pareto undominated. Further, for Maximin $0^+$-Cloning can be solved in linear time. However, no election is 1-manipulable by cloning with respect to Maximin.*

*Proof.* Clearly, if all voters prefer some alternative $a$ to $c$, then after any cloning the Maximin score of $c$ will be 0, whereas the Maximin score of at least one alternative is positive, and hence $c$ cannot win. On the other hand, if $c$ is undominated, its Maximin score is at least 1. Now, we can use the construction described above to reduce the Maximin score of any other candidate to 1, thus making $c$ a winner.

Our algorithm for Maximin $0^+$-Cloning relies on the observation that the only way to change the Maximin score of a candidate is to clone her, thereby reducing her score. Now, suppose that $Sc_M(c) = s$. We will argue that we have a "yes"-instance of $0^+$-Cloning if and only if our budget allows us to introduce $\lceil \frac{n}{s} \rceil$ clones for each candidate $a$ such that $Sc_M(a) > s$; clearly, this condition can be checked in linear time.

Indeed, for each candidate $a$ whose Maximin score exceeds $s$ we can do the following. We create $\lceil \frac{n}{s} \rceil$ clones of $a$, divide the voters in $s$ groups, where the size of the first $s - 1$ groups is $\lceil \frac{n}{s} \rceil$, and the last group consists of the remaining $t \leq \lceil \frac{n}{s} \rceil$ voters. For each of the first $s - 1$ groups, we ask the voters to rank the clones according to an $\lceil \frac{n}{s} \rceil$-cyclic election; the voters in the last group vote as the first $t$ voters in an $\lceil \frac{n}{s} \rceil$-cyclic election. Clearly, in each group, for any $i = 1, \ldots, \lceil \frac{n}{s} \rceil$ there is at most one voter who ranks the $i$-th clone above the $(i - 1)$-st clone. Thus, in the resulting election, the Maximin score of any clone of $a$ is at most $s$, and therefore $c$ is a winner of that election.

For the converse direction, we need to show that if we create less than $\lceil \frac{n}{s} \rceil$ clones of $a$, the Maximin score of at least one of them will exceed $s$. Indeed, suppose that we create $t < \lceil \frac{n}{s} \rceil$ clones of $a$; denote these clones by $a^{(1)}, \ldots, a^{(t)}$. Given an arbitrary preference profile over these clones, consider a directed graph whose vertices are the clones and there is an edge from $a^{(i)}$ to $a^{(j)}$ if at least $n - s$ voters prefer $a^{(i)}$ to $a^{(j)}$. Note that if the Maximin score of each clone is at most $s$, each vertex of this graph must have an incoming edge. In particular, this means that our graph cannot be acyclic. We will now argue that each cycle in this graph has length at least $\lceil \frac{n}{s} \rceil$; clearly, this implies that the graph contains at least $\lceil \frac{n}{s} \rceil$ vertices, a contradiction.

To see this, suppose that there is a cycle of length $r < \lceil \frac{n}{s} \rceil$. Relabel the clones along this cycle as $a^{(1)}, \ldots, a^{(r)}$, i.e., assume that for each $i = 1, \ldots, r$ at least $n-s$ voters prefer $a^{(i)}$ to $a^{(i+1)}$ (where $a^{(r+1)} = a^{(1)}$). By induction on $i$ it is easy to see that for each $i = 1, \ldots, r-1$, there are at least $n - si$ voters whose preference order $\succ$ satisfies $a^{(1)} \succ \ldots \succ a^{(i+1)}$. For $i = r-1$, this implies that at least $n - s(r-1) > s$ voters prefer $a^{(1)}$ to $a^{(r)}$, a contradiction with the assumption that there is an edge from $a^{(r)}$ to $a^{(1)}$. This establishes that our algorithm for $0^+$-CLONING is correct.

Finally, it is easy to see that no election is 1-manipulable with respect to Maximin. Indeed, the only way to change a candidate's Maximin score is to clone him. However, after the cloning, all voters may order all clones in the same way, in which case the most popular clone will have the same Maximin score as the original alternative. $\qquad\square$

It is not clear if one can strengthen the result of Theorem 4.7 to $q$-manipulability for $q \in (0, 1)$. This amounts to the following question: Suppose that for a fixed $n$ we randomly draw $n$ permutations of $\{1, \ldots, k\}$. Let $P(n, k)$ be the probability that for each $i \in [k]$ there is a $j \in [k]$ such that $j$ precedes $i$ in at least $n - 1$ permutations. Is it the case that, as $k \to \infty$, the probability $P(n, k)$ approaches 1?

Our computations show that this is unlikely to be the case.[3] For $(n, k) = (5, 20)$ there was only one success out of $10^6$ random trials and only three for $(n, k) = (5, 50)$. For both $(n, k) = (7, 20)$ and $(n, k) = (7, 50)$ not a single random trial out of $10^6$ trials was successful.

This means that, even if Maximin is $q$-manipulable for some fixed $q > 0$, the number of clones needed would be astronomical.

## 5. Three Rules for Which Cloning May Be Difficult

We now consider Borda, $k$-Approval, and Copeland rules, for which cloning-related problems are significantly more difficult.

### 5.1 Borda Rule

For the Borda rule, the necessary and sufficient condition for the existence of a $0^+$-successful manipulation by cloning is the same as for Maximin: the manipulator's favorite alternative has to be Pareto undominated. However, Borda and Maximin exhibit different behavior with respect to 1-manipulability. Moreover, from the point of view of finding a minimum-cost cloning, Borda appears to be harder to deal with than Maximin.

**Theorem 5.1.** *An election is $0^+$-manipulable by cloning with respect to Borda if and only if the manipulator's preferred candidate $c$ does not win, but is Pareto undominated. Moreover, UC $0^+$-CLONING for Borda can be solved in linear time.*

*Proof.* Note first that if we create $k$ clones of an alternative $a$ that was ranked in position $j$ in some order $R_i$, then $R_i$'s contribution to the scores of all alternatives ranked above $a$ increases by $k - 1$, $R_i$'s contribution to the scores of all alternatives ranked below $a$ does not change, and, finally, the top-ranked clone of $a$ receives $k-1$ more points from $R_i$ than $a$ used to receive. We will now use this observation to prove the theorem.

---

3. We are grateful to Danny Chang for his help in performing these experiments.

Suppose first that all voters prefer some alternative $a$ to $c$. Note that this implies that $Sc_B(a) > Sc_B(c)$. If we create $k$ clones of some alternative $x \neq c, a$, for each preference ordering $R_i$ there are three possibilities:

1. In $R_i$, $x$ is ranked above both $a$ and $c$. Then cloning $x$ does not change the Borda scores of $a$ and $c$.

2. In $R_i$, $x$ is ranked below $a$ and above $c$. Then cloning $x$ increases the Borda score of $a$ by $k-1$ and does not change the score of $c$.

3. In $R_i$, $x$ is ranked below both $a$ and $c$. Then cloning $x$ increases the Borda scores of both $a$ and $c$ by $k-1$.

Thus, any single act of cloning an alternative in $A \setminus \{c, a\}$ (and hence any combination of them) does not reduce the gap between the scores of $a$ and $c$. Further, if we clone $a$ and/or $c$, in every ordering $R_i$ each clone of $a$ is ranked above each clone of $c$, and hence has a higher Borda score. Thus, $c$ cannot be made a winner by cloning.

For the converse direction, let $C = \{a \in A \mid Sc_B(a) > Sc_B(c)\}$. For each $a \in C$, let $n_a$ be the number of voters that prefer $c$ to $a$, i.e., $n_a = W(c, a)$. Let $s_a$ denote the score difference between $a$ and $c$, i.e., $s_a = Sc_B(a) - Sc_B(c)$. Now, set $k = \max_{a \in C} \lceil \frac{s_a}{n_a} \rceil$, and create $k+1$ clones of $c$. Consider the preference profile in which all voters rank all clones of $c$ in the same way. Let $c^*$ be the top-ranked clone of $c$. Observe that the Borda score of $c^*$ in the new profile exceeds the original Borda score of $c$ by $kn$, where $n$ is the total number of voters. Now, consider some alternative $a \in C$. For each order $R_i$ in which $a$ was ranked above $c$, $R_i$'s contribution to the score of $a$ increased by $k$, and for each order $R_i$ in which $a$ was ranked below $c$, $R_i$'s contribution to the score of $a$ remained the same. Thus, $a$'s score has increased by $k(n - n_a)$. Hence, using $Sc'_B(a)$ and $Sc'_B(c^*)$ to denote the scores of $a$ and $c^*$ after cloning, by our choice of $k$ we have $Sc'_B(a) - Sc'_B(c^*) = Sc_B(a) - Sc_B(c) - kn_a \leq 0$. We conclude that in the resulting preference profile the Borda score of $c^*$ is at least as high as that of any other alternative, i.e., our cloning is $0^+$-successful.

We will now argue that our input constitutes a "yes"-instance of UC $0^+$-Cloning if and only if $\max_{a \in C} \lceil \frac{s_a}{n_a} \rceil + 1 \leq B$, where $B$ is the cloning budget, i.e., the manipulation constructed above is optimal in the UC model. Indeed, consider an alternative $a \in C$ that maximizes the expression $\frac{s_a}{n_a}$. By creating $t+1$ clones of some alternative $d \neq c, a$ we can increase the distance between $c$ and $a$ in the orders in which $d$ is ranked below $c$ but above $a$ by $t$, and thus reduce the gap between $a$ and $c$ by $t$. Obviously, there are at most $n_a$ such orders. For any other order $R_i$, cloning $d$ may increase $R_i$'s contribution to the score of $a$, or affect $R_i$'s contributions to scores of $a$ and $c$ in the same way. Thus, creating $t+1$ clones of an alternative $d \in A \setminus \{a, c\}$ can contribute at most $tn_a$ to closing the gap between $c$ and $a$. A similar argument applies to cloning $c$ or $a$, showing that the contribution from $t+1$ clones of either of these alternatives is also bounded by $tn_a$. Hence, we have to create at least $\lceil \frac{s_a}{n_a} \rceil + 1$ clones. $\square$

Is it possible to strengthen Theorem 5.1 to $q$-manipulability for some constant $q \in (0, 1]$? It turns out that for Borda these questions are significantly more difficult than for the rules we have considered so far.

We will first consider the situation where the manipulator is restricted to cloning her favorite candidate. We remark that this special case of our model is very natural: For instance, a party may be able to nominate several candidates, but is not in a position to force other parties to do so. In this case, for an even number of voters, we can characterize the elections that can be 1-manipulated by cloning with respect to Borda.

Let $c$ be the manipulator's preferred candidate. We will show that, by cloning $c$, we can deal with the candidates who currently have a higher Borda score than $c$, as long as they lose to $c$ in a pairwise election. However, we have to be careful to ensure that $c$'s Borda score remains higher than that of the candidates who beat $c$ in a pairwise election. Formally, let

$$A^+ = \{a \in A \setminus \{c\} \mid S_B(a) > S_B(c)\}, \quad A^- = \{a \in A \setminus \{c\} \mid S_B(a) \leq S_B(c)\}.$$

Let $s_a = |S_B(a) - S_B(c)|$ for each $a \in A \setminus \{c\}$, and set

$$n_a = \begin{cases} W(c,a) - W(a,c) & \text{if } a \in A^+, \\ W(a,c) - W(c,a) & \text{if } a \in A^-. \end{cases}$$

Finally, let $r^+ = +\infty$ if $n_a \leq 0$ for some $a \in A^+$ and $r^+ = \max\{\frac{2s_a}{n_a} \mid a \in A^+\}$ otherwise, and let $r^- = \min\{\frac{2s_a}{n_a} \mid a \in A^-, n_a > 0\}$. We are now ready to state our criterion.

**Theorem 5.2.** *An election with an even number of voters can be 1-manipulated with respect to Borda by cloning the manipulator's preferred candidate $c$ if and only if $c$ does not win and $\lceil r^+ \rceil \leq \lfloor r^- \rfloor$.*

*Proof.* Let $n$ denote the number of voters. Suppose that $\lceil r^+ \rceil > \lfloor r^- \rfloor$. Consider any cloning that involves $c$ only. Suppose it results in $k$ clones of $c$, which we denote by $c^{(1)}, \ldots, c^{(k)}$. Let $s$ denote the original Borda score of $c$. To show that this cloning is not 1-successful, it suffices to describe an ordering of the clones that results in all clones of $c$ losing the election.

Consider a profile where the first $\frac{n}{2}$ voters rank the clones as $c^{(1)} \succ \cdots \succ c^{(k)}$, while the remaining $\frac{n}{2}$ voters rank the clones in the opposite order. Clearly, the Borda score of any clone is $s + \frac{n}{2}(k-1)$. We will now consider two cases.

**Case 1 ($r^+ = +\infty$).** This means that there is an alternative $a$ such that $n_a \leq 0$, i.e., $a$ is preferred to $c$ by at least $\frac{n}{2}$ voters and has a higher Borda score than $c$. Our cloning increases $a$'s score by at least $\frac{n}{2}(k-1)$, so its final Borda score is at least $s_B(a) + \frac{n}{2}(k-1) > s + \frac{n}{2}(k-1)$. Thus, after cloning, all clones of $c$ will still have lower scores than $a$, i.e., the cloning is not 1-successful.

**Case 2 ($r^+ < +\infty$).** In this case, there exists a candidate $a \in A^+$ such that $n_a > 0$ and $r^+ = \frac{2s_a}{n_a}$. Consider a candidate $b \in A^-$ such that $r^- = \frac{2s_b}{n_a}$; we have $n_b > 0$. The condition $\lceil r^+ \rceil > \lfloor r^- \rfloor$ can be rewritten as $\lceil \frac{2s_a}{n_a} \rceil > \lfloor \frac{2s_b}{n_b} \rfloor$. After the cloning, $a$'s Borda score is $s_B(a) + \frac{n-n_a}{2}(k-1)$, and $b$'s Borda score is $s_B(b) + \frac{n+n_b}{2}(k-1)$. Thus, for $c$ to be a winner, $k$ must satisfy

$$s + \frac{n(k-1)}{2} \geq s_B(a) + \frac{n-n_a}{2}(k-1); \quad s + \frac{n(k-1)}{2} \geq s_B(b) + \frac{n+n_b}{2}(k-1).$$

Hence, we obtain

$$s_a = s_B(a) - s \leq \frac{n_a}{2}(k-1), \quad s_b = s - s_B(b) \geq \frac{n_b}{2}(k-1).$$

Since $k$ is integer, this implies $r^+ = \lceil \frac{2s_a}{n_a} \rceil \leq k - 1 \leq \lfloor \frac{2s_b}{n_b} \rfloor = r^-$, a contradiction.

For the opposite direction, we will show that the ordering of the clones constructed above is the worst possible for the manipulator. Formally, suppose that we generate $k$ clones of $c$ for some $k \geq 1$. Consider a voter who gives $j$ points to $c$. It is not hard to see that this voter gives $j + (j+1) + \cdots + (j+k-1) = kj + \frac{k(k-1)}{2}$ points to all clones of $c$. Thus, the total Borda score of all clones is equal to $ks + n\frac{k(k-1)}{2}$, where $s$ is the Borda score of $c$ prior to cloning. It follows that the Borda score of at least one clone is $s + \frac{n}{2}(k-1)$ or higher.

Since $\lceil r^+ \rceil \leq \lfloor r^- \rfloor$, we have $r^+ < +\infty$. Set $k = 1 + \lceil r^+ \rceil$, and consider an arbitrary alternative $a \in A^+$. As $r^+ < +\infty$, we have $n_a > 0$. Therefore, $a$ gains $\frac{n-n_a}{2}(k-1)$ points from cloning. Now, it is not hard to see that $k \geq 1 + \frac{2s_a}{n_a}$ implies $s + \frac{n}{2}(k-1) \geq S_B(a) + \frac{n-n_a}{2}(k-1)$, which means that after the cloning some clone of $c$ beats $a$.

To finish the proof, consider a candidate $b \in A^-$. If $n_b \leq 0$, after the cloning the score of $b$ is at most $S_B(b) + n\frac{k-1}{2} \leq s + n\frac{k-1}{2}$, so $b$ still loses to or ties with some clone of $c$. On the other hand, if $n_b > 0$, the cloning increases the score of $b$ by $\frac{n+n_b}{2}(k-1)$ points. We have

$$k - 1 = \lceil r^+ \rceil \leq \lfloor r^- \rfloor \leq \left\lfloor \frac{2s_b}{n_b} \right\rfloor \leq \frac{2s_b}{n_b}.$$

Therefore, after the cloning, the score of $b$ is

$$S_B(b) + \frac{n+n_b}{2}(k-1) \leq S_B(b) + \frac{n}{2}(k-1) + s_b = s + n\frac{k-1}{2},$$

i.e., $b$'s Borda score does not exceed that of some clone of $c$. $\square$

It is not hard to see that the second part of the proof (the "if" direction) works for the odd number of voters as well. However, for the "only if" direction, the argument does not go through. We can, however, prove a slightly weaker necessary condition. Define $\hat{r}^+ = +\infty$ if $n_a \leq 0$ for some $a \in A^+$ and $\hat{r}^+ = \max\{\frac{2s_a-1}{n_a} \mid a \in A^+\}$ otherwise, and let $\hat{r}^- = \min\{\frac{2s_a+1}{n_a} \mid a \in A^-, n_a > 0\}$.

**Proposition 5.3.** *If an election with an odd number of voters can be 1-manipulated with respect to Borda by cloning the manipulator's preferred candidate $c$, then $\lceil \hat{r}^+ \rceil \leq \lfloor \hat{r}^- \rfloor$.*

We relegate the proof of Proposition 5.3 to Appendix A. We remark that it is not clear if the converse direction of Proposition 5.3 holds: the condition $\lceil \hat{r}^+ \rceil \leq \lfloor \hat{r}^- \rfloor$ is weaker than $\lceil r^+ \rceil \leq \lfloor r^- \rfloor$, and we were unable to prove that it is sufficient for 1-manipulability with respect to Borda.

The proof of Theorem 5.2 indicates which orderings of the clones are most problematic for the manipulator: these are the orderings that grant each clone roughly the same number of points. But this is exactly the expected outcome if the orderings are generated uniformly at random! Thus, our proof shows that for Borda a manipulator who clones $c$ only should be prepared for the worst-case scenario.

Note, however, that we are not limited to cloning $c$. Indeed, cloning other candidates might be more useful with respect to 1-manipulability. For example, suppose that $c$ is Pareto undominated, and, moreover, the original preference profile contains a candidate $c'$ that is ranked right under $c$ by all voters (one can think of this candidate as an "inferior clone" of $c$; however, we emphasize that it is present in the original profile). Then one can show that by cloning $c'$ sufficiently many times we can make $c$ a winner with probability 1. However, cloning $c$ itself does not have the same effect if the voters order the clones randomly or adversarially to the manipulator. This is illustrated by the following example.

**Example 5.4.** Let us consider the following Borda election. The set of candidates is $C = \{a, b, c, d\}$, and there are four voters, whose preference orders are:

$$R_1 : a \succ c \succ b \succ d \qquad Sc_B(a) = 9$$
$$R_2 : a \succ c \succ b \succ d \qquad Sc_B(b) = 4$$
$$R_3 : a \succ c \succ b \succ d \qquad Sc_B(c) = 8$$
$$R_4 : d \succ c \succ b \succ a \qquad Sc_B(d) = 3$$

The winner here is $a$ with 9 points. However, replacing $b$ with three clones $b_1, b_2, b_3$ is a 1-successful manipulation in favor of $c$ since the new score of $a$ is 15, while the new score of $c$ is 16, no matter how the clones are ordered. At the same time, we cannot make $c$ a winner with probability 1 by cloning him alone. Indeed, if we split $c$ into $k + 1$ clones, and each voter orders the clones uniformly at random, the expected score of each clone of $c$ is $4(2 + \frac{k}{2}) = 8 + 2k$ (and, since the number of voters is even, the proof of Theorem 5.2 gives an explicit ordering of the clones in which each clone gets $8 + 2k$ points), whereas $a$'s score is $9 + 3k$.

This shows that, in general, we may need to clone several candidates that are placed between $c$ and its "competitors" in a large number of votes, and determining the right candidates to clone might be difficult. Indeed, it is not clear if a 1-successful manipulation for Borda can be found in polynomial time; answering this question is a challenging open problem.

A related question that is not answered by Theorem 5.1 is the complexity of $0^+$-CLONING (and, more generally, $q$-CLONING for $q \in \{0^+\} \cup (0, 1]$) in the general cost model. Note that there is a certain similarity between this problem and that of 1-manipulability: in both cases, it may be suboptimal to clone $c$. Indeed, for general costs, we can prove that $q$-CLONING is NP-hard for any $q$.

**Theorem 5.5.** *For Borda, $q$-CLONING in the general cost model is NP-hard for any rational $q \in (0, 1]$ as well as for $q = 0^+$. Moreover, this is the case even if $p(i, j) \in \{0, 1, \infty\}$ for all $i \in [m], j \in \mathbb{Z}^+$.*

*Proof.* We provide a reduction from X3C. Let $(G, \mathcal{S})$ be an instance of X3C, where $G = \{g_1, \ldots, g_{3K}\}$ is the ground set and $\mathcal{S} = \{S_1, \ldots, S_M\}$ is a family of 3-element subsets of $G$. Let $N = 3K$. We construct an instance of our problem as follows. Let $T = \{t_1, \ldots, t_M\}$. We let the set of alternatives be $A = G \cup T \cup \{c, u, w\}$, where $c$ is the manipulator's preferred alternative. For each $i = 1, \ldots, M$, we create two voters with preference orders $R_{2i-1}$ and $R_{2i}$. The preference order $R_{2i-1}$ is given by

$$R_{2i-1} : G \setminus S_i \succ c \succ t_i \succ S_i \succ u \succ w \succ T \setminus \{t_i\},$$

while the preference order $R_{2i}$ is given by

$$R_{2i} : w \succ S_i \succ u \succ c \succ G \setminus S_i \succ T.$$

Further, we require that the candidates in $G$ are ranked in the opposite order in $R_{2i}$ and $R_{2i-1}$, i.e., for any $j, \ell = 1, \ldots, M$ we have $g_j \succ_{2i} g_\ell$ if and only if $g_\ell \succ_{2i-1} g_j$. Finally, there is one voter whose preference order is given by

$$R_{2M+1} : G \succ c \succ T \succ u \succ w.$$

The cloning costs are defined as follows. For each $i = 1, \ldots, M$, producing one additional clone of $t_i$ costs 1, and producing further clones of $t_i$ costs 0. Cloning any other alternative costs $+\infty$. Finally, we set $B = K$.

It is easy to verify that from each pair of votes $(R_{2i-1}, R_{2i})$

1. $c$ gets $(M + 5) + (N - 3 + M) = 2M + N + 2$ points,

2. each $g_j$ gets $(M + N + 2) + M = 2M + N + 2$ points (this remains true if $g_j \in S_i$),

3. each $t_j$ gets at most $(M + 4) + (M - 1) = 2M + 3$ points,

4. $u$ gets $M + (M + N - 2) = 2M + N - 2$ points, and

5. $w$ gets $(M - 1) + (M + N + 2) = 2M + N + 1$ points.

Thus, because of the last voter, the overall Borda score of each alternative in $G$ exceeds that of $c$ by at least 1 and at most $N$, while the score of any other alternative is lower than that of $c$.

Note that if we create $N + 1$ clones of $t_i$, which we can do at cost 1, we ensure that $c$'s score is at least as high as that of the alternatives in $S_i$, irrespective of how the voters order the clones. However, cloning $t_i$ does not change the difference in scores between $c$ and the alternatives in $G \setminus S_i$. Within our budget, we can clone $K$ alternatives from $T$, $N + 1$ times each. Thus, any cover of $G$ of size $K = N/3$, i.e., any exact cover, can be transformed into a 1-successful cloning manipulation for our instance.

Conversely, consider any cloning manipulation of cost at most $K$. Suppose that it clones the alternatives $t_{i_1}, \ldots, t_{i_s}$, $s \leq K$. If $\{S_{i_1}, \ldots, S_{i_s}\}$ is not a cover of $G$, there is some element $g_i \in G$ that is not covered by $\cup_{j=1}^{s} S_{i_j}$. This means that our cloning manipulation did not change the difference in scores between $g_i$ and $c$, i.e., $g_i$ still has a higher Borda score than $c$. Hence, this cloning is not $0^+$-successful. Thus, any $0^+$-successful cloning of cost at most $K$ corresponds to a cover of $G$.

Now, consider any $q \in (0, 1]$. If our input instance of X3C is a "yes"-instance, the election that we have constructed admits a 1-successful cloning, which is also a $q$-successful cloning (as well as a $0^+$-successful cloning). On the other hand, if we start with a "no"-instance of X3C, our election does not admit a $0^+$-successful cloning, and hence a $q$-successful cloning. Thus, the proof is complete. □

Observe that the cost function used in the proof of Theorem 5.5 is very natural: some candidates cannot be cloned at all, while for others, there is a certain upfront cost associated with creating the first clone (e.g., researching the platform and/or identity of the candidate), but all subsequent clones can be created for free.

## 5.2 *k*-Approval

We will now demonstrate that there is a family of scoring rules for which deciding whether a given election is $0^+$-manipulable is computationally hard. Specifically, this is the case for *k*-Approval for any $k \geq 2$. Our proof gives a reduction from the problem DOMINATING SET, defined below.

**Definition 5.6.** *An instance of the* DOMINATING SET *problem is a triple* $(V, E, s)$, *where* $(V, E)$ *is an undirected graph and* $s$ *is an integer. We ask if there is a subset* $W$ *of* $V$ *such that (a)* $|W| \leq s$ *and (b) for each* $v \in V$ *either* $v \in W$ *or* $v$ *is connected to a vertex in* $W$.

**Theorem 5.7.** *For the* $k$-*Approval rule, it is* NP-*hard to decide whether a given election is* $0^+$-*manipulable by cloning.*

*Proof.* First, we remark that one can always make an alternative $a$ a $k$-Approval winner with nonzero probability as long as $a$ is ranked first by at least one voter. Indeed, we can simply clone all other alternatives sufficiently many times (e.g., $kn$ times, where $n$ is the number of voters). This ensures that there exists a preference profile over the resulting set of alternatives in which each candidate other than $a$ gets at most one $k$-Approval point. However, this condition is not necessary. Indeed, a candidate can be a $k$-Approval winner even if no voter ranks him first. Thus, in our hardness reduction, we ask whether we can help a candidate that is never ranked first.

Consider an instance $(V, E, s)$ of DOMINATING SET, where $V = \{v_1, \ldots, v_t\}$. We can assume without loss of generality that the graph $(V, E)$ has no isolated vertices and $s < t$. We will construct an election based on $(V, E, s)$ as follows. There is a candidate $v_i$ for each vertex of the graph, a candidate $c$ whom we would like to make a winner, an additional candidate $w$, and a set $D$ of dummy candidates. The exact number of dummy candidates, which will be polynomial in the size of $(V, E)$, will become clear after we describe the set of voters. For each voter, we only specify which of the nondummy candidates she ranks in the top $k$ positions. Clearly, the order of candidates ranked in positions $k + 1$ or lower does not affect the outcome of the election. Further, we assume that each dummy candidate is ranked among the top $k$ positions by exactly one voter. (Thus, the number of dummy candidates is bounded from above by $k$ times the number of voters.)

For each $i = 1, \ldots, t$, the $i$-th voter places $v_i$ first and ranks $c$ in position $k$. Each of the next $4t^2 - t - s$ voters places $w$ first and ranks $c$ in position $k$. Further, for each (undirected) edge $(v_i, v_j) \in E$, there are $2s$ voters that rank $v_i$ first and $v_j$ second and $2s$ voters that rank $v_j$ first and $v_i$ second.

| $v_1$ | $\cdots$ | $v_t$ | $w$ | $\cdots$ | $w$ | $\cdots$ | $v_i$ | $\cdots$ | $v_i$ | $v_j$ | $\cdots$ | $v_j$ | $\cdots$ |
|---|---|---|---|---|---|---|---|---|---|---|---|---|---|
| $\vdots$ | $\cdots$ | $\vdots$ | $\vdots$ | $\cdots$ | $\vdots$ | $\vdots$ | $v_j$ | $\cdots$ | $v_j$ | $v_i$ | $\cdots$ | $v_i$ | $\vdots$ |
| $c$ | $\cdots$ | $c$ | $c$ | $\cdots$ | $c$ | $\cdots$ | $\vdots$ | $\cdots$ | $\vdots$ | $\vdots$ | $\cdots$ | $\vdots$ | $\vdots$ |
| | $t$ | | | $4t^2 - t - s$ | | | | $2s$ | | | $2s$ | | |

Observe that based on the votes constructed so far, the score of $c$ is $4t^2 - s$, the score of $w$ is $4t^2 - t - s$, and the score of each $v_i$, $i = 1, \ldots, t$, is at most $4s(t - 1) + 1$. We now add polynomially many voters, each of which ranks $w$ or some candidate from $V$ first (and some dummy candidates in positions $2, \ldots, k$), so that $w$ gets $4t^2 - 2s$ points in total and each candidate in $V$ gets exactly $4t^2$ points in total (note that this is possible since $t > s$). Clearly, the number of the voters constructed at this stage is polynomially bounded in our input size.

We claim that the constructed election is $0^+$-manipulable if and only if the graph $(V, E)$ has a dominating set of size at most $s$. Indeed, suppose first that $(V, E)$ has a dominating set $W$ of size at most $s$. We clone each candidate from $W$ exactly $k + 2s(k-1)$ times. Now, consider the following preference profile over the cloned alternatives. Split the clones into $2s + 1$ groups, so that the first group has size $k$ and the remaining groups have size $k - 1$. For each $v_i \in W$, we order the clones as follows:

1. The $i$-th voter ranks the first group of $v_i$'s clones in the first $k$ positions.

2. For each $j$ such that $(v_j, v_i) \in E$, the $\ell$-th voter among the $2s$ voters that used to rank $v_j$ first and $v_i$ second now ranks the $(\ell + 1)$-st group of $v_i$'s clones in positions $2, \ldots, k$.

3. All other voters order the clones arbitrarily.

Under this preference profile, $c$ gets $4t^2 - 2s$ points: he has been "pushed out" of the top $k$ slots in $s$ out of the first $t$ votes, but did not lose any other points. The clone of each candidate $v_i \in W$ gets at most $4t^2 - 2s$ points, too. Indeed, fix some $v_i \in W$. By our assumption, the graph $(V, E)$ has no isolated vertices. Thus, there is some vertex $v_j$ connected to $v_i$, and so there are $2s$ voters that (before cloning) rank $v_j$ in the first position and $v_i$ in the second position. Let us denote this set of voters by $E_{ji}$. Now, consider the profile after cloning. The first $k$ clones of $v_i$ are not ranked in the first $k$ positions by any voter in $E_{ji}$, and thus each of them receives at most $4t^2 - 2s$ points. Now consider the $\ell$-th group of $v_i$'s clones, $\ell = 2, \ldots, 2s + 1$: these clones are ranked in the top $k$ positions by exactly one voter in $E_{ji}$, and by none of the first $t$ voters. Thus, these clones' scores are at most $4t^2 - (2s - 1) - 1 = 4t^2 - 2s$. We conclude that the score of each clone of $v_i$ is at most $4t^2 - 2s$.

Now, consider any candidate $v_j \in V \setminus W$. Since $W$ is a dominating set for $(V, E)$, there exists a $v_i \in W$ such that $(v_i, v_j) \in E$. Since the $2s$ voters that used to rank $v_i$ first and $v_j$ second now rank some clones of $v_i$ in the top $k$ positions, $v_j$ is "pushed out" of the top $k$ positions in at least $2s$ votes. Consequently, its $k$-Approval score is at most $4t^2 - 2s$. As the $k$-Approval score of $w$ was $4t^2 - 2s$ prior to cloning and was not affected by cloning, we conclude that after cloning $c$ is a winner of the election with nonzero probability.

Conversely, suppose that we can make $c$ a winner with nonzero probability by cloning some of the candidates. Observe first that we could not have cloned $w$. Indeed, if we clone $w$, $c$ is "pushed out" of the top $k$ positions in $4t^2 - t - s$ votes, so her score is at most $t$. On the other hand, in the original preferences, each $v_i \in V$ is ranked first at least $4t$ times. Therefore, to make $c$ a winner after cloning $w$, we would have to clone each of $v_i \in V$. However, if we do that, $c$'s $k$-Approval score goes down to 0, so she cannot be a winner. We conclude that $w$ is not cloned. As $w$ is not ranked in positions $2, \ldots, k$ in any of the votes, and therefore cannot be "pushed out", this means that $w$'s final score is $4t^2 - 2s$, and therefore $c$'s final score must be at least $4t^2 - 2s$. This, in turn, means that we can clone at most $s$ candidates in $V$, as cloning any such candidate reduces $c$'s score by 1. On the other hand, we need to reduce the score of each $v_i \in V$ by at least $s$. This can be done either by cloning $v_i$ or by cloning some alternative that is ranked first by some voter that ranks $v_i$ second. This means that each $v_i \in V$ either is cloned or has a neighbor in $(V, E)$ that has been cloned, i.e., the set of cloned alternatives forms a dominating set of $(V, E)$. As we have argued that the size of this set is at most $s$, this completes the reduction. $\square$

We can also show that it is NP-hard to decide whether an election is 1-manipulable with respect to $k$-Approval.

**Theorem 5.8.** *For any given $k \geq 2$, it is* NP*-hard to decide whether a given election is 1-manipulable by cloning with respect to $k$-Approval.*

*Proof.* We will first present the proof for $k = 2$, and then show how to generalize it to all integer $k \geq 2$. The reduction is from X3C. Let an instance $(G, \mathcal{S})$ of X3C be given by a ground set $G = \{g_1, \ldots, g_{3K}\}$ and a family $\mathcal{S} = \{S_1, \ldots, S_M\}$ of subsets of $G$, where for each $i = 1, \ldots, M$ we write $g_{i_1}, g_{i_2}, g_{i_3}$ to denote the members of $S_i$. We can assume without loss of generality that $M > K > 3$. Given such an instance, we construct an election as follows. The set of alternatives is $A = G \cup T \cup \{c, w\} \cup D$, where $T = \{t_1, \ldots, t_M\}$, and $D$ is the set of dummy candidates. In what follows, we place the dummy candidates so that each of them appears at most once in the first two rows.

We will now specify the first two positions in each vote. For $i = 1, \ldots, M$, the $i$-th voter ranks $t_i$ first and $c$ second. Each of the next $M^2$ voters ranks $w$ first and a dummy alternative second. Then we also have $M$ groups of $3M$ voters, where in the $i$-th group (a) each voter ranks $t_i$ first, and (b) each of the candidates $g_{i_1}$, $g_{i_2}$, and $g_{i_3}$ is ranked second by exactly $M$ voters. Further, for each $i \in G$, we add $N_i = M(M + 1 - |\{j \mid g_i \in S_j\}|)$ voters that rank $g_i$ first and a dummy candidate second. Finally, we add $N = M^2 - M + K$ voters who rank $c$ first and a dummy candidate second.

| $t_1$ | $\cdots$ | $t_M$ | $w$ | $\cdots$ | $w$ | $\cdots$ | $t_i$ | $t_i$ | $t_i$ | $\cdots$ | $g_i$ | $\cdots$ | $g_i$ | $\cdots$ | $c$ | $\cdots$ | $c$ |
|---|---|---|---|---|---|---|---|---|---|---|---|---|---|---|---|---|---|
| $c$ | $\cdots$ | $c$ | $\cdots$ | $\cdots$ | $\cdots$ | $\cdots$ | $g_{i_1}$ | $g_{i_2}$ | $g_{i_3}$ | $\cdots$ | $\cdots$ | $\cdots$ | $\cdots$ | $\cdots$ | $\cdots$ | $\cdots$ | $\cdots$ |
| | $M$ | | | $M^2$ | | | | $3$ | | | | $N_i$ | | | | $N$ | |

In this profile, the score of each $t_i$ is $3M + 1$, the score of $c$ is $M^2 + K$, the score of $w$ is $M^2$, and the score of each $g_i$ is $M^2 + M$. The scores of dummy candidates are equal to 1, and their number is clearly polynomial in $M$. As $K < M$, the set of winners in this profile is $G$, and $c$ is $M - K$ points behind. Now, if we clone some candidate $g_i \in G$, the voters may still rank all clones of $g_i$ in the same order, so no manipulation that relies on cloning candidates in $G$ is 1-successful. Therefore, to bridge the gap between $c$ and the candidates in $G$, we need to clone some of the candidates in $T$. However, we cannot clone more than $K$ of them, since otherwise the score of $c$ will fall below the score of $w$ (while we can clone $w$ as well, the voters may still order all clones of $w$ in the same way, in which case the top-ranked clone of $w$ still gets $M^2$ points). On the other hand, for each $g_i \in G$ we need to clone at least one $t_j \in T$ such that $g_i \in S_j$, since otherwise the score of $g_i$ will not go down. Thus, a 1-successful manipulation corresponds to a cover of $G$ of size at most $K$, i.e., an exact cover. Conversely, cloning a set of candidates $T' \subseteq T$ such that $\{S_j \mid t_j \in T'\}$ is an exact cover of $G$ is a 1-successful cloning.

For $k > 2$, this construction can be modified as follows. In each vote, we insert a group of new $k - 2$ dummy candidates between the first and the second position (so that in each vote the candidate that used to be ranked second is now ranked in position $k$ and each dummy candidate is ranked in top $k$ positions exactly once). It is easy to see that the proof goes through without change. □

## 5.3 Copeland

To analyze cloning under the Copeland rule, we need some additinal definitions. For an election $E$ with a set of candidates $A$, its *pairwise majority graph* is a directed graph $(A, X)$, where $X$ contains an edge from $a$ to $b$ if more than half of the voters prefer $a$ to $b$; we say that $a$ *beats* $b$ if $(a, b) \in X$. If exactly half of the voters prefer $a$ to $b$, we say that $a$ and $b$ are *tied* (this does not mean that their Copeland scores are equal). For any $a \in A$, we denote by $U(a)$, $D(a)$ and $T(a)$ the sets of all alternatives that beat $a$, are beaten by $a$, or are tied with $a$, respectively.

For an odd number of voters, the graph $(A, X)$ is a *tournament*, i.e., for each pair $(a, b) \in A^2$, $a \neq b$, we have either $(a, b) \in X$ or $(b, a) \in X$. In this case, we can make use of a well-known tournament solution concept of Uncovered Set (Miller, 1977; Fishburn, 1977; Laslier, 1997), defined as follows. Given a tournament $(A, X)$, a candidate $a$ is said to *cover* another candidate $b$ if $a$ beats $b$ as well as every other candidate beaten by $b$. The *Uncovered Set* of $(A, X)$ is the set of all candidates not covered by other candidates.

It turns out that if the number of voters is odd, the Uncovered Set coincides with the set of candidates that can be made Copeland winners by cloning. In contrast with our previous characterization results, this holds for all values of $q \in (0, 1] \cup \{0^+\}$.

**Theorem 5.9.** *For any $q \in (0, 1] \cup \{0^+\}$, a Copeland election $E$ with an odd number of voters is $q$-manipulable by cloning if and only if the manipulator's preferred candidate $c$ does not win, but is in the Uncovered Set of the pairwise majority graph of $E$.*

*Proof.* Consider an election $E$ with a set of $m$ alternatives $A$ and an odd number of voters. Let $(A, X)$ be its pairwise majority graph. Suppose first that $c$ is covered by some $a \in A$. In this case $Sc_C(c) < Sc_C(a)$. Creating $k$ clones of an alternative $x$ increases by $k - 1$ the score of each alternative that beats $x$ and decreases by $k - 1$ the score of each alternative that is beaten by $x$. Hence, the gap between $c$ and $a$ cannot be reduced by cloning a third alternative. Moreover, if we replace $c$ with $k$ clones, the score of each clone of $c$ will be at most $Sc_C(c) + k - 1$, while the score of $a$ becomes $Sc_C(a) + k - 1$. Similarly, if we replace $a$ with $k$ clones, the scores of each clone of $a$ will be at least $Sc_C(a) - (k - 1)$, while the score of $c$ becomes $Sc_C(c) - (k - 1)$. This shows that cloning $c$ or $a$ would not help either. Thus, no matter which alternatives we clone, we cannot close the gap between $a$ and $c$ (or its highest-scoring clone).

Conversely, suppose that $c$ is in the Uncovered Set. Since the number of candidates is odd, we have $T(c) = \emptyset$, and since $c$ is uncovered, we have $U(c) \neq A \setminus \{c\}$. Therefore, $D(c) \neq \emptyset$. We will proceed in two stages. First, we will secure that either $c$ or its clone has a higher Copeland score than any alternative from $D(c)$, and then we will take care of the alternatives from $U(c)$. At stage one we create $2m + 1$ clones of $c$. This lowers the score of each alternative in $D(c)$ by $2m$ and raises the score of each alternative in $U(c)$ by $2m$. A simple counting argument shows that, for any ordering of the clones by the voters, there exists a clone of $c$ whose score is greater or equal to the original score of $c$; denote this clone

551

by $c'$. For each alternative $x$, let $Sc'_C(x)$ denote $x$'s score after stage one. For each $x \in D(c)$ we have

$$Sc'_C(c') \geq Sc_C(c) \geq -(m-1) > Sc_C(x) - 2m = Sc'_C(x),$$

i.e., $c'$ has a higher Copeland score than $x$. On the other hand, for any $x \in U(c)$ we have

$$Sc'_C(c') \geq Sc_C(c) \geq -(m-1) \geq Sc_C(x) - (2m-2) = Sc'_C(x) - (4m-2).$$

At stage two we create $4m + 1$ clones of each alternative in $D(c)$. This increases the score of $c'$ by $4m|D(c)|$. Further, for any $a \in D(c)$, the score of any clone of $a$ constructed at this stage exceeds $Sc'_C(a)$ by at most $4m|D(c)|$. Thus, at this stage, $c'$ has a higher Copeland score than any of the newly-generated clones. Finally, since $c$ was not covered, no candidate in $U(c)$ beats all candidates in $D(c)$; in fact, any $a \in U(c)$ is beaten by some $b \in D(c)$. Thus, the last step increases the score of each candidate in $U(c)$ by at most $4m(|D(c)| - 1) - 4m = 4m(|D(c)| - 2)$. It follows that $c'$ now has a higher Copeland score than any candidate that is not a clone of $c$. $\square$

For elections with an even number of voters, the situation is significantly more complicated. The notion of Uncovered Set can be extended to pairwise majority graphs of arbitrary elections in a natural way (see, e.g., Brandt & Fischer, 2007): we say that $u$ covers $c$ if $u$ beats $c$ and all alternatives beaten by $c$, and, in addition, $c$ loses to all alternatives that beat $u$. In particular, this means that $u$ *does not* cover $c$ if it is beaten by some alternative that is tied with $c$. This definition generalizes the one for the odd number of voters. However, for an even number of voters, the condition that $c$ is in the Uncovered Set turns out to be necessary, but not sufficient for manipulability by cloning.

**Proposition 5.10.** *In a Copeland election $E$ with an even number of voters in which the manipulator's preferred candidate $c$ is covered, $c$ cannot be made a winner by cloning.*

*Proof.* Suppose that $c$ is covered by $u$. The Copeland score of $c$ is given by $|D(c)| - |U(c)|$. Now, $u$ beats all alternatives in $D(c)$, and any alternative that beats $u$ is necessarily in $U(c)$. Thus, $u$ has a strictly higher Copeland score than $c$. Now, suppose we create $k + 1$ clones of some $a \in A$. This can increase the score of $c$ (or one of its clones) only if $a \in D(c) \cup \{c\}$; however, then the score of $u$ also increases by $k$. On the other hand, this can decrease the score of $u$ only if $a$ beats $u$. However, in this case the score of $c$ also decreases by $k$. We conclude that $c$ cannot be made a winner by cloning. $\square$

However, the converse is not true, as illustrated by the following example.

**Example 5.11.** Consider an election with $A = \{a, b, c, u, w\}$. Suppose that $a$ beats $u$, $u$ beats $b$, $b$ beats $a$, $w$ beats $a$, and $w$ beat $c$, and any other pair of candidates is tied. Note that by McGarvey's theorem (1953) there are voters' preferences that produce this pairwise majority graph. In this election, $c$ is undominated. Indeed, while he is beaten by $u$ and $w$, he is tied with an alternative that beats $u$, namely, $a$, and with an alternative that beats $w$, namely, $b$. However, $c$'s score is negative, and will remain negative after any cloning. On the other hand, by a counting argument, in any election where some alternative has a negative Copeland score, at least one other alternative has a positive Copeland score. Hence, $c$ cannot be made a winner by cloning.

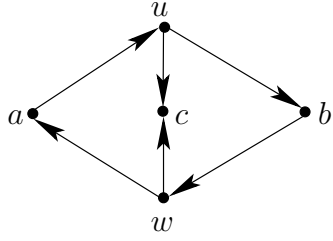

Figure 1: Candidate $c$ is undominated, but cannot be made a winner by cloning

Instead, we can characterize $0^+$-manipulable profiles in terms of the properties of the induced (bipartite) subgraph of $(A, X)$ whose vertices are, on the one hand, the candidates that are tied with $c$, and, on the other hand, the candidates that beat both $c$ and all candidates beaten by $c$. However, it is not clear if this characterization leads to a polynomial-time algorithm for detecting such profiles.

Specifically, for a profile in which $c$ is not covered to be $0^+$-manipulable by cloning in favor of $c$, the following must hold. Let $Y = T(c)$ and let $Z$ be the set of all candidates that beat both $c$ and all candidates beaten by $c$. We associate with each candidate $z \in Z$ a number $s_z = Sc_C(z) - Sc_C(c)$. Our goal is to assign a nonnegative integer $q(y)$ to every $y \in Y$ so that for each $z \in Z$ we have

$$\sum_{(y,z) \in X} q(y) - \sum_{(z,y) \in X} q(y) \geq s_z. \tag{1}$$

Indeed, if this linear program has an integer nonnegative solution $(q^*(y))_{y \in Y}$, we can replace each $y \in Y$ with $q^*(y) + 1$ clones. This will lower the score of each $z \in Z$ by $\sum_{(y,z) \in X} q^*(y) - \sum_{(z,y) \in X} q^*(y)$ and thus ensure that $c$'s Copeland score is at least as high as that of any candidate in $Z$. We can then take care of the rest of the candidates by cloning $c$ (and asking all voters to order all clones of $c$ in the same way) and the candidates in $D(c)$, as in the proof of Theorem 5.9. Thus, condition (1) is sufficient for $0^+$-manipulability. (We remark, however, that additional constraints may be needed for 1-manipulability, since cloning $c$ may have a stronger effect on candidates in $Z$ than on the "average" clone of $c$.)

Conversely, if such an assignment is impossible, there is no way to ensure that all candidates in $Z$ have a lower Copeland score than $c$. Indeed, cloning $c$ or the candidates in $D(c)$ will not help, as any such candidate is also beaten by any candidate in $Z$. On the other hand, cloning a candidate in $U(c)$ will harm $c$ (or its clones) at least as much as it will harm the candidates in $Z$. Thus, the only way to close the gap between $c$ and the candidates in $Z$ is to clone candidates in $T(c)$, as captured by our integer linear program. Note also that if the linear program (1) does not admit an integer solution, this remains to be the case if we clone some of the candidates. Indeed, cloning candidates in $A \setminus (Y \cup Z)$ does not change (1). Cloning a candidate $z \in Z$ replaces an existing constraint with several identical ones. Finally, if the program obtained by cloning a candidate $y \in Y$ has a feasible solution, it can be easily transformed into a feasible solution for the original program.

Once we introduce costs, optimal cloning becomes hard even for elections with an odd number of voters and even in the UC model.

**Theorem 5.12.** *For Copeland,* UC $q$-CLONING *is* NP-*hard for each* $q \in \{0^+\} \cup (0, 1]$.

*Proof.* We give a reduction from X3C. Let $I = (G, \mathcal{S})$ be our input instance, where $G = \{g_1, \ldots, g_{3K}\}$ is the ground set and $\mathcal{S} = \{S_1, \ldots, S_M\}$ is a family of 3-element subsets of $G$. It will be convenient to assume that $M \geq 2K$; we can always achieve this by duplicating some sets in $\mathcal{S}$. We will construct an instance of Copeland elections with a preferred candidate $c$ such that if $I$ is a "yes"-instance of X3C, then there is a 1-successful cloning that introduces at most $K$ clones, and, furthermore, if $I$ is a "no"-instance of X3C, then there does not exist a $0^+$-successful cloning that introduces at most $K$ clones.

The set of candidates for our election will be $A = G \cup F \cup D \cup \{c\}$, where $c$ is the manipulator's preferred candidate, $F = \{f_1, \ldots, f_M\}$, and $D = \{d_1, \ldots, d_{3M+2}\}$. The cloning budget is set to $K$. We describe the voters by providing the outcomes of all head-to-head contests between candidates in $A$. By McGarvey's theorem (McGarvey, 1953), any such tournament can be realized as the majority relation of a certain preference profile, which can be computed in polynomial time. The results of head-to-head contests between the candidates in $G \cup F \cup \{c\}$ will be as follows:

1. $c$ beats each $f_j \in F$.

2. Each $g_i \in G$ beats $c$.

3. For each $g_i \in G$ and each $S_j \in \mathcal{S}$, $g_i$ beats $f_j$ if and only if $g_i \notin S_j$.

4. All the remaining contests within $G \cup F \cup \{c\}$ result in a tie.

If we limit ourselves to candidates in $G \cup F \cup \{c\}$, we have the following scores: (a) $c$ has $M - 3K$ points, (b) each $g_i \in G$ has between 1 and $M + 1$ points, and (c) each $f_j \in F$ has at most 0 points.

Further, we set the results of head-to-head contests between candidates in $G \cup F \cup \{c\}$ and those in $D$ as follows:

5. $c$ beats $2M + 2K \leq 3M$ candidates in $\{d_1, \ldots, d_{3M}\}$.

6. Each $g_i \in G$ beats exactly as many candidates in $\{d_1, \ldots, d_{3M}\}$ as to have score $3M - K + 1$.

7. All the other contests between candidates in $G \cup F \cup \{c\}$ and those in $D$ result in a tie.

Finally, we set the contests within $D$ as follows: Each $d_i \in \{d_1, \ldots, d_{3M}\}$ loses to $d_{3M+1}$ and to $d_{3M+2}$, and all the remaining contests within $D$ result in a tie. Let $N = 3M$. In the resulting election we have the following scores:

1. $c$ has $N - K$ points.

2. Each $g_i \in G$ has $N - K + 1$ points.

3. $d_{N+1}$ and $d_{N+2}$ have $N$ points each.

4. Every other candidate has at most 0 points.

Consider some nonnegative integer $k$. Replacing a single candidate with $k+1$ clones increases other candidates' scores by at most $k$, and the score of each clone can differ from that of the cloned candidate by at most $k$. As a result, the only candidates that can be winners after a cloning manipulation of cost $K$ are those in $G \cup \{c, d_{N+1}, d_{N+2}\}$. Thus, in the following discussion we will consider the scores of these candidates only.

Introducing a single clone of some $f_j \in F$ increases by 1 the scores of $c$ and of each $g_i \notin S_j$. Thus, if there is a set $J \subseteq \{1, \ldots, M\}$ such that $|J| = K$ and $\cup_{i \in J} S_i = G$, then introducing a single clone of each candidate in $\{f_i \mid i \in J\}$ ensures that $c$ is a winner of the election. This does not depend on how the voters order the introduced clones, i.e., this cloning strategy is 1-successful.

For the converse direction, we will now argue that if there is a $0^+$-successful cloning of cost at most $K$, then $I$ is a "yes"-instance of X3C. Let us consider such a $0^+$-successful cloning. We have to introduce $K$ clones, because each additional clone can increase $c$'s score by at most 1, $c$ trails $d_{N+1}$ and $d_{N+2}$ by $K$ points, and after any cloning some clones of $d_{N+1}$ and $d_{N+2}$ will have at least $N$ points. Consequently, we cannot clone any of the candidates in $D$: cloning any $d_i \in D \setminus \{d_{N+1}, d_{N+2}\}$ increases the scores of $d_{N+1}$ and $d_{N+2}$, and cloning $d_{N+1}$ or $d_{N+2}$ does not increase $c$'s score. Cloning any $g_i \in G$ also does not increase $c$'s score, and thus we can only clone members of $F \cup \{c\}$. We will now argue that the cloned members of $F$ correspond to a cover of $G$ (note that this implies that there is exactly $K$ of them, and each of them is cloned exactly once). Indeed, suppose otherwise, and let $g_i$ be an element of $G$ that is not covered by the union of the sets that correspond to cloned members of $F$. Then our cloning manipulation increases the score of $g_i$ by $K$, as each new clone contributes to the increase of $g_i$'s score. On the other hand, any cloning that is within our budget can increase $c$'s score by at most $K$, so $c$ still trails $g_i$, a contradiction with the assumption that our cloning manipulation is $0^+$-successful. This completes the proof. □

## 6. Related Work

We will now review several lines of research that are related to our work. We start by discussing the relevant work that originates in the social choice community (Section 6.1), and then move on to the computational study of voting problems. In Section 6.2 we provide a detailed comparison between cloning and the classic problem of control by adding candidates (Bartholdi et al., 1992), and in Section 6.3 we discuss other related work in computational social choice.

### 6.1 Cloning in Social Choice Literature

The first study of cloning was undertaken by Tideman (1987). He started by defining which subsets of the alternative set $A$ are clones for a given profile. Specifically, he defined a proper subset $S$ of $A$ that contains at least two alternatives to be a *clone-set* if no voter ranks any candidate outside of $S$ in between the members of $S$ or as tied with some member of $S$. This definition reflects the idea that each voter finds the candidates in $S$ similar. Now, every profile $\mathcal{R}$ on $A$ defines a set of clone-sets $\mathcal{C}(\mathcal{R}) \subseteq 2^A$. Tideman defines a voting rule to be *independent of clones* if and only if the following two conditions are met when clones are on the ballot:

1. A candidate that is a member of a set of clones wins if and only if some member of that set of clones wins after a member of the set is eliminated from the ballot.

2. A candidate that is not a member of a set of clones wins if and only if that candidate wins after any clone is eliminated from the ballot.

Tideman considered a number of well-known voting rules, and discovered that among these rules STV was the only one that satisfied his criterion. However, STV does not satisfy many other important criteria for voting rules, such as Condorcet consistency or monotonicity. Thus, Tideman proposed a new voting rule, the "ranked pairs rule," that was both Condorcet consistent and independent of clones in all but a small fraction of settings. Subsequently, Zavist and Tideman (1989) proposed a modification of this rule that is completely independent of clones. Later it was shown that some other voting rules, such as Schulze's rule (2003), are also independent of clones. Tideman considered resistance to cloning to be an important normative requirement for voting rules, and the method of ranked pairs proposed by him and Zavist showed how this condition may be satisfied.

Another notion related to cloning is that of composition consistency, due to Laffond et al. (1996). It is defined for tournament solution concepts (i.e., social choice correspondences that take tournaments as inputs). For a given alternative set $A$, let $\mathcal{T}(A)$ denote the set of all tournaments on $A$. A non-empty subset $C$ of $A$ is a *component* of $T \in \mathcal{T}(A)$ if for any $c, c' \in C$ and for any $a \in A \setminus C$ it holds that $(a, c) \in T$ if and only if $(a, c') \in T$; a component $C$ is said to be *nontrivial* if $1 < |C| < |A|$. A tournament $T$ is said to be *composed* if it has at least one nontrivial component. Components of a tournament are natural counterparts of clone-sets of a preference profile.

Let $A_1, \ldots, A_K$ be $K$ disjoint finite sets of alternatives, and let $T_1, \ldots, T_K$ be $K$ tournaments such that for each $k = 1, \ldots, K$ it holds that $T_k \in \mathcal{T}(A_k)$. Moreover, let $T^*$ be a tournament in $\mathcal{T}([K])$, where $[K] = \{1, 2, \ldots, K\}$. The *composition product* of $T^*$ by tournaments $T_1, \ldots, T_K$ is the tournament $T = \Pi(T^*; T_1, \ldots, T_K)$ defined on $A = A_1 \cup \ldots \cup A_K$ as follows. For each $k, k' \in [K]$ and each $(a, b) \in A_k \times A_{k'}$ the tournament $T$ contains the pair $(a, b)$ if and only if (i) $k \neq k'$ and $(k, k') \in T^*$ or (ii) $k = k'$ and $(a, b) \in T_k$.

A tournament solution concept $\phi$ is said to be *composition-consistent* if for each composition product $T = \Pi(T^*; T_1, \ldots, T_K)$ in $\mathcal{T}(A)$ it holds that $\phi(T) = \bigcup \{\phi(T_k) \mid k \in \phi(T^*)\}$. This property may be illustrated in the following way. Suppose that there are $K$ different projects that a given society can choose to implement. Further, each project $A_k$, $k \in [K]$, has $n_k$ different variants. Composition consistency guarantees that we will choose the best variant of the best project irrespectively of whether we use a two-stage procedure that first chooses the best project and then its best variant, or simply set up a single tournament in order to choose among all variants of all projects.

Laffond et al. (1996) and Laslier (1996) suggested a similar construction for social choice correspondences (which are referred to as voting rules in this paper). Given a set $X$, let $\mathbf{R}_n(X)$ be the set of all $n$-voter profiles on $X$. Let $A_1, \ldots, A_K$ be $K$ disjoint finite sets of alternatives, and consider $K$ preference profiles $\mathcal{R}^1, \ldots, \mathcal{R}^K$, where $\mathcal{R}^k = (R_1^k, \ldots, R_n^k)$ is a profile in $\mathbf{R}_n(A_k)$ for $k \in [K]$. Let $\mathcal{R}^* = (R_1^*, \ldots, R_n^*) \in \mathbf{R}_n([K])$. The *composition product* $\Pi(\mathcal{R}^*; \mathcal{R}^1, \ldots, \mathcal{R}^K)$ of $\mathcal{R}^*$ by profiles $\mathcal{R}^1, \ldots, \mathcal{R}^K$ is the $n$-voter profile $\mathcal{R} = (R_1, \ldots, R_n)$ defined on $A = A_1 \cup \ldots \cup A_K$ as follows. For each $k, k' \in [K]$ and each $(a, b) \in A_k \times A_{k'}$ we set $a \, R_i \, b$ if and only if (i) $k \neq k'$ and $k \, R_i^* \, k'$ or (ii) $k = k'$ and $a \, R_i^k \, b$. A social choice correspondence $\phi$ is said to be *composition-consistent* if for any $n > 0$ and any composition product $\mathcal{R} = \Pi(\mathcal{R}^*; \mathcal{R}^1, \ldots, \mathcal{R}^K)$ in $\mathbf{R}_n(A)$ it holds that $\phi(\mathcal{R}) = \bigcup\{\phi(\mathcal{R}^k) \mid k \in \phi(\mathcal{R}^*)\}$.

Laffond et al. (1996) and Laslier (1996) proved that a number of tournament solution concepts such as the Banks Set, the Uncovered Set, the Tournament Equilibrium Set (TEQ), and the Minimal Covering Set are composition-consistent, but several other tournament solution concepts and social choice correspondences such as the Top Cycle, the Slater rule, the Copeland rule, and all scoring rules are not composition-consistent.

Laslier (2000) also introduced the notion of *cloning consistency*. A social choice correspondence $\phi$ is said to be *cloning-consistent* if for any $n > 0$ and any composition product $\mathcal{R} = \Pi(\mathcal{R}^*; \mathcal{R}^1, \ldots, \mathcal{R}^K)$ in $\mathbf{R}_n(A)$, it holds that $\phi(\mathcal{R}) = \bigcup\{A_k \mid k \in \phi(\mathcal{R}^*)\}$. This requirement says that if one clone of an alternative is winning, then all clones of that alternative must win as well. This property is useful when the set of alternatives is fuzzy, but—as the author himself acknowledged—terrible if the number of alternatives is fixed and clearly defined.

The concept of a composed profile is not necessarily useful for manipulation by cloning, but is very relevant to what one might call *decloning*. The idea of decloning is to reveal whether or not a profile (or, a tournament) could have been obtained as a result of cloning and, if possible, identify the underlying composition product. Recently, decloning proved to be a useful preprocessing tool for dealing with voting rules that have a computationally hard winner determination problem (Conitzer, 2006; Betzler, Fellows, Guo, Niedermeier, & Rosamond, 2009; Brandt, Brill, & Seedig, 2011). The extended abstract of the current paper, which was presented at the Twenty-Fourth AAAI Conference on Artificial Intelligence (AAAI-2010), initiated a complexity-theoretic study of decloning in voting; this topic was further investigated by Elkind, Faliszewski, and Slinko (2011). However, decloning deserves careful study on its own and thus has been omitted from this paper.

The concept of resistance to cloning appeared also in the context of study of self-selectivity of social choice functions. Koray and Slinko (2008) discovered that self-selectivity is a stronger requirement than resistance to cloning, even if deletion of candidates is viewed as a special form of cloning (one that replaces a candidate with zero clones). This partially explains why the universally self-selective functions are necessarily dictatorial, as discovered earlier by Koray (2008). Koray and Slinko (2008) circumvent this impossibility result by relaxing the property of self-selectivity: they require the social choice function to select itself only among other "reasonable" social choice functions. The concept of being reasonable involves a social choice correspondence (for example, the one that selects all Pareto optimal alternatives), and it is essential that this social choice correspondence is resistant to cloning of essential alternatives (for the definition of an essential alternative see Koray & Slinko, 2008).

## 6.2 Comparison of Cloning and Other Models of Adding Candidates

A problem that is closely related to cloning is that of election control. In general, this term refers to manipulating the result of an election by changing its structure (e.g., by either adding or deleting candidates or voters). The computational study of election control was initiated by Bartholdi et al. (1992) who, among other issues, considered constructive control by adding candidates (CCAC). In their model, we are given a set of registered candidates, a set of spoiler candidates, and a set of voters, with preferences over both the registered candidates and the spoiler candidates (however, before we take any action, only the registered candidates participate in the election). The task is to decide if it is possible to select a subset of spoiler candidates so that when these candidates are registered, the preferred candidate becomes a winner. Subsequently, Faliszewski, Hemaspaandra, Hemaspaandra, and Rothe (2009) refined this model by introducing a bound on the number of candidates that can be added.

Formally, we will use the following definition of the CCAC problem, which is based on the one given by Faliszewski et al. (2009).

**Definition 6.1.** *Let $\mathcal{F}$ be a voting rule. In the* constructive control by adding candidates *problem ($\mathcal{F}$-CCAC) we are given an election $(C \cup A, \mathcal{R})$, where $C \cap A = \emptyset$, a designated candidate $p \in C$, and a nonnegative integer $t$. We ask if there is a set $A' \subseteq A$ of size at most $t$ such that $p$ is the unique $\mathcal{F}$-winner of the election $(C \cup A', \mathcal{R})$.*[4]

In the definition above, the set $C$ corresponds to already registered candidates and the set $A$ is the set of "spoiler" candidates that the manipulator can introduce into the election.

Bartholdi, Tovey, and Trick considered only two rules, Plurality and Condorcet's rule (i.e., the rule that selects a Condorcet winner if one exists and no winners otherwise), and focused on standard worst-case complexity results, classifying control problems as either NP-complete or in P. Many researchers followed up on their work by studying various other voting rules (Erdélyi, Nowak, & Rothe, 2009b; Faliszewski et al., 2009; Faliszewski, Hemaspaandra, & Hemaspaandra, 2011) and various other settings (Liu, Feng, Zhu, & Luan, 2009; Betzler & Uhlmann, 2009; Faliszewski et al., 2011), of which perhaps most prominent are destructive control of Hemaspaandra, Hemaspaandra, and Rothe (2007) and control in multi-winner elections of Meir, Procaccia, Rosenschein, and Zohar (2008). We point the reader to the recent survey of Faliszewski, Hemaspaandra, and Hemaspaandra (2010) for more details on election control.

While $q$-CLONING (and, in particular, UC $q$-CLONING) and CCAC control are similar in that both of them deal with adding new candidates, neither of these problems is a special case of the other. Indeed, they place different restrictions on the candidates to be added and their positions in the votes. Specifically, in $q$-CLONING the new candidates must be clones of existing candidates, but (especially in $0^+$-CLONING) we have some freedom as to how to arrange the new candidates in the votes. In contrast, in CCAC control problems, the spoiler candidates need not be adjacent to each other in all votes, but the order of all the candidates in each vote is predetermined.

A somewhat different model of adding candidates has been recently proposed by Chevaleyre, Lang, Maudet, and Monnot (2010). In this paper, the authors consider the following scenario. An election is happening over a period of time and candidates may still join in. At a given point, we know all the candidates that have registered by then and the voters' preferences over those candidates. Each voter may place new candidates arbitrarily in her vote. Given that at most $k$ new candidates may still appear, which of the already registered ones still have a chance of winning? (Note that, as in the case of cloning, the addition of new candidates may benefit some of the original candidates and hurt some of the others.) Chevaleyre et al. (2010), and—in a very recent follow-up work—Xia, Lang, and Monnot (2011), give computational complexity results for the problem of finding the possible (co)winners when new alternatives join (PcWNA). Their work differs from ours in that they do not require the new candidates to be clones of preexisting ones (in particular, they do not require the new candidates to be ranked consistently, consecutively by all the voters), and differs from control by adding candidates in that they allow introducing arbitrary new candidates (as opposed to introducing already-ranked-by-voters candidates from a predefined set of spoilers). Formally, their model is a special case of the possible winner problem, introduced by Konczak and Lang (2005), and further studied by many other researchers (Xia & Conitzer, 2011; Betzler & Dorn, 2010; Bachrach, Betzler, & Faliszewski, 2010; Baumeister & Rothe, 2010).

---

4. Unique-winner model is standard for control problems.

Table 1 provides a comparison of the complexity of control by adding candidates (CCAC control), possible co-winner determination when new alternatives can join (PcWNA), and $q$-Cloning for $q \in \{0^+, 1\}$ in the UC model (using the UC model is analogous to counting the number of new candidates in the CCAC control problems). To fill the first column of Table 1, we provide complexity results for CCAC control for Plurality with Runoff, $k$-approval, Borda, and Veto, which were missing from the existing literature. Since these results are tangential to the topic of our paper, we relegate them to Appendix B.

Table 1 shows that cloning and PcWNA are computationally incomparable (assuming $P \neq NP$). For example, for 2-Approval $0^+$-Cloning is NP-hard, whereas PcWNA is in P. On the other hand, for Maximin $0^+$-Cloning is in P, while PcWNA is NP-complete.

In contrast, Table 1 appears to indicate that CCAC control is harder than cloning. These results are not entirely surprising: we can construct contrived instances of CCAC control by placing the spoiler candidates in any way we like, so as to facilitate computational hardness proofs. One may therefore conjecture that UC $0^+$-Cloning is always easier than CCAC control. However, it turns out that this is not the case.

**Theorem 6.2.** *There is a voting rule for which constructive control by adding candidates is in* P*, but* UC $0^+$*-Cloning is* NP*-hard.*

The proof of Theorem 6.2 is given in Appendix B; while the voting rule constructed in this proof is highly artificial, it demonstrates that UC $0^+$-Cloning cannot be reduced to CCAC control (unless $P = NP$).

To wrap up our discussion of control by adding candidates, we mention an interesting twist to the standard model of election control that was recently studied by Faliszewski, Hemaspaandra, Hemaspaandra, and Rothe (2011) and Brandt, Brill, Hemaspaandra, and Hemaspaandra (2010). In these papers, the authors study the complexity of control (as well as manipulation and bribery) for single-peaked electorates. Their main finding is that many control problems that are known to be NP-hard for unrestricted preferences (and most control problems belong to this category) turn out to be solvable in polynomial time when the preferences are single-peaked. In comparison, cloning is computationally feasible for many rules even without assuming special properties of the electorate. As a side note, we mention that cloning a candidate may destroy single-peakedness of an election: if each voter ranks the clones uniformly at random, the resulting ranking of the clones is unlikely to be single-peaked. The problem of "collapsing" the minimum number of clones in order to make a given election single-peaked is studied in detail by Elkind et al. (2011).

| Voting rule | CCAC control | PcWNA | $0^+$-Cloning | 1-Cloning |
|---|---|---|---|---|
| Plurality | NPC (Bartholdi et al., 1992) | P (Betzler & Dorn, 2010) | P | — |
| Maximin | NPC (Faliszewski et al., 2011) | NPC (Xia et al., 2011) | P | — |
| Plurality w/Runoff | NPC | P (Xia et al., 2011) | P | — |
| Veto | NPC | P (Betzler & Dorn, 2010) | P | P |
| Borda | NPC | P (Chevaleyre et al., 2010) | P | ? |
| 2-Approval | NPC | P (Chevaleyre et al., 2010) | NP-hard | NP-hard |
| $k$-Approval, $k \geq 3$ | NPC | NPC (Chevaleyre et al., 2010) | NP-hard | NP-hard |
| Copeland | NPC (Faliszewski et al., 2009) | ? | NP-hard | NP-hard |

Table 1: The complexity of control via adding candidates (CCAC), of possible co-winner determination when new alternatives can join (PcWNA), and of $q$-Cloning in the UC model for $q \in \{0^+, 1\}$. Note that for Plurality, Plurality with Runoff and Maximin 1-successful cloning is impossible. All the results on CCAC control are in the unique-winner model (though some of them are also proved in the non-unique winner model), whereas we work in the non-unique winner model. The PcWNA results for Plurality and Veto follow directly from the more general results of Betzler and Dorn (2010) for the possible winner problem. Xia et al. (2011) also give an NP-completeness result for a variant of Copeland rule known as Copeland[0], where the score of a candidate is simply the number of head-to-head contests that this candidate wins.

### 6.3 Cloning and Campaign Management

To a large extent, our work on cloning is motivated by applications of cloning in campaign management. However, campaign management can be understood in multiple other ways as well. In particular, the issue of campaign management in voting has been previously studied from the computational perspective by Elkind, Faliszewski, and Slinko (2009), who introduced the problem of swap bribery. In their model, each voter is associated with a certain cost function, which describes how difficult it is to make local changes to this voter's preferences. The goal of the campaign manager is to ensure that a given candidate becomes a winner at the smallest possible cost. While this problem turns out to be NP-hard for almost all voting rules, some of its special cases admit polynomial-time solutions. Further, if one focuses on a variant of swap bribery where one is only allowed to shift forward the preferred candidate, it is possible to find effective (approximation) algorithms (Elkind et al., 2009; Elkind & Faliszewski, 2010; Schlotter, Elkind, & Faliszewski, 2011). Going in a different research direction, Dorn and Schlotter (2010) provide parameterized complexity study of swap bribery. Of course, the standard model of bribery (Faliszewski, Hemaspaandra, & Hemaspaandra, 2009), where one can pay a voter to change his vote arbitrarily, can also be interpreted in the context of campaign management.

Also, the probabilistic model put forward in this paper, and, in particular, our definition of $q$-successful cloning is similar in spirit to the model of Erdélyi, Fernau, Goldsmith, Mattei, Raible, and Rothe (2009a), where voters are bribed to increase their probabilities of voting in favor of a particular alternative.

Finally, the problem of cloning is particularly relevant in open, anonymous environments, such as the Internet. In such settings, a problem closely related to cloning candidates is that of cloning *voters*. Specifically, in an anonymous environment an agent might be capable of creating several instances of itself and vote multiple times. Voting rules that are resistant to this kind of manipulation are called false-name-proof; they were studied by Conitzer (2008). A variant of this framework in which, similarly to our general model, creating new identities is costly was subsequently considered by Wagman and Conitzer (2008).

## 7. Conclusions

We have provided a formal model of manipulating elections by cloning, characterized $0^+$-manipulable and 1-manipulable profiles for many well-known voting rules, and explored the complexity of finding a minimum-cost cloning manipulation. The grouping of voting rules according to their susceptibility to manipulation differs from most standard classifications of voting rules: e.g., scoring rules behave very differently from each other, and Maximin is more similar to Plurality than to Copeland. Future research directions include designing approximation algorithms for the minimum-cost cloning under voting rules for which this problem is known to be NP-hard, and extending our results to other voting rules.

### Acknowledgments

We would like to thank the anonymous JAIR referees for their very useful feedback. Edith Elkind is supported by NRF (Singapore) Research Fellowship (NRF-RF2009-08) and an NTU start-up grant. Piotr Faliszewski is supported by AGH University of Technology

Grant no. 11.11.120.865, by Polish Ministry of Science and Higher Education grant N-N206-378637, and by Foundation for Polish Science's program Homing/Powroty. Arkadii Slinko is supported by the Faculty of Science Research and Development Fund grant 3624495/9844.

## Appendix A. Cloning the Manipulator's Preferred Candidate under Borda: Odd Number of Voters

In this section, we present the proof of Proposition 5.3. (We use the notation introduced in the paragraph preceding Theorem 5.2.)

**Proposition 5.3.** If an election with an odd number of voters can be 1-manipulated with respect to Borda by cloning the manipulator's preferred candidate $c$, then $\lceil \hat{r}^+ \rceil \leq \lfloor \hat{r}^- \rfloor$.

*Proof.* Let $n$ be the number of voters. Suppose that $\lceil \hat{r}^+ \rceil > \lfloor \hat{r}^- \rfloor$. Consider any cloning that involves $c$ only. Suppose it results in $k$ clones of $c$, which we denote by $c^{(1)}, \ldots, c^{(k)}$. Let $s$ denote the original Borda score of $c$. To show that this cloning is not 1-successful, it suffices to describe an ordering of the clones that results in $c$ losing the election. Since for $n = 1$ no cloning is successful, we can assume without loss of generality that $n \geq 3$.

Suppose first that $k$ is odd. Consider the profile[5] where the first voter ranks the clones as

$$c^{(k)} \succ c^{(k-2)} \succ \ldots \succ c^{(1)} \succ c^{(k-1)} \succ c^{(k-3)} \succ \ldots \succ c^{(2)},$$

the second voter ranks the clones as

$$c^{(k-1)} \succ c^{(k-3)} \succ \ldots \succ c^{(2)} \succ c^{(k)} \succ c^{(k-2)} \succ \ldots \succ c^{(1)},$$

the third voter ranks the clones as

$$c^{(1)} \succ \ldots \succ c^{(k)},$$

and the remaining voters are split into $\frac{n-3}{2}$ pairs, where in each pair the first voter ranks the clones as $c^{(k)} \succ \ldots \succ c^{(1)}$, while the second voter ranks the clones as $c^{(1)} \succ \ldots \succ c^{(k)}$.

It is not hard to see that in this profile the Borda score of each clone is $s + \frac{n(k-1)}{2}$. Indeed, let us first consider $c^{(k)}$. The first voter ranks $k-1$ other clones below $c^{(k)}$, the second voter ranks $\frac{k-1}{2}$ other clones below $c^{(k)}$, and the third voter ranks no other clones below $c^{(k)}$, i.e., $c^{(k)}$ gets $\frac{3(k-1)}{2}$ additional points from the first 3 voters. Also, he gets $k-1$ additional point from each pair of the remaining voters, i.e., $\frac{(n-3)(k-1)}{2}$ additional points. Thus, his Borda score is $s + \frac{n(k-1)}{2}$. A similar calculation shows that the Borda score of $c^{(k-1)}$ is $s + \frac{k-3}{2} + (k-1) + 1 + \frac{(n-3)(k-1)}{2} = s + \frac{n(k-1)}{2}$. Now, if we compare two consecutive odd-numbered clones, i.e., $c^{(j)}$ and $c^{(j-2)}$ for $j$ odd, we can see that $c^{(j)}$ is ranked just above $c^{(j-2)}$ in the first two votes, and two positions below $c^{(j)}$ in the third vote, so they get the same number of extra points from the first three voters. Since any two candidates get the same number of votes from the last $n-3$ voters, it follows that $c^{(j)}$ and $c^{(j-2)}$ have the same Borda scores. The same argument applies to any pair of consecutive even-numbered clones. Hence, a simple inductive argument shows that the Borda score of each clone is $s + \frac{n(k-1)}{2}$.

---

5. We are grateful to Dima Shiryaev for suggesting this construction.

Now, if $k$ is even, we set $k' = k - 1$, and rank the first $k'$ clones using the construction for odd $k$ given above. We then place $c^{(k)}$ above all other clones in the first $\frac{n-1}{2}$ votes and below all other clones in the remaining $\frac{n+1}{2}$ votes. Clearly, the score of any clone $c^{(j)}$, $j < k$, is

$$s + \frac{n(k-2)}{2} + \frac{n+1}{2} = s + \frac{n(k-1)+1}{2} = s + \left\lceil \frac{n(k-1)}{2} \right\rceil;$$

the last equality holds since $n$ is odd and $k$ is even. Moreover, the score of $c^{(k)}$ is $s + (k-1)\frac{n-1}{2} \le s + \lceil \frac{n(k-1)}{2} \rceil$. Thus, for all values of $k$ the Borda score of each clone is at most $s + \lceil \frac{n(k-1)}{2} \rceil$.

To show that in this profile $c$ does not win, we will consider two cases.

**Case 1 ($\hat{r}^+ = +\infty$).** Then there is an alternative $a$ that is preferred to $c$ by at least $\frac{n+1}{2}$ voters and has a higher Borda score than $c$. Our cloning increases $a$'s score by at least $\frac{n+1}{2}(k-1)$, so its final Borda score is at least $s_B(a) + \frac{n+1}{2}(k-1) > s + \lceil \frac{n(k-1)}{2} \rceil$. Thus, after cloning, all clones of $c$ will still have lower scores than $a$, i.e., the cloning is not 1-successful.

**Case 2 ($\hat{r}^+ < +\infty$).** In this case, there exist candidates $a \in A^+$, $b \in A^-$ such that $\lceil \frac{2s_a - 1}{n_a} \rceil > \lfloor \frac{2s_b + 1}{n_b} \rfloor$. After the cloning, $a$'s Borda score is $s_B(a) + \frac{n - n_a}{2}(k-1)$, and $b$'s Borda score is $s_B(b) + \frac{n + n_b}{2}(k-1)$. Thus, for $c$ to be the winner, $k$ must satisfy

$$s + \left\lceil \frac{n(k-1)}{2} \right\rceil \ge s_B(a) + \frac{n - n_a}{2}(k-1); \quad s + \left\lceil \frac{n(k-1)}{2} \right\rceil \ge s_B(b) + \frac{n + n_b}{2}(k-1).$$

We have $\lceil \frac{n(k-1)}{2} \rceil \le \frac{n(k-1)+1}{2}$. Thus, by rewriting the above inequalities, we obtain

$$s_a = s_B(a) - s \le \frac{n_a}{2}(k-1) + \frac{1}{2}, \quad s_b = s - s_B(b) \ge \frac{n_b}{2}(k-1) - \frac{1}{2}.$$

Since $k$ is an integer, this implies

$$\hat{r}^+ = \left\lceil \frac{2s_a - 1}{n_a} \right\rceil \le k - 1 \le \left\lfloor \frac{2s_b + 1}{n_b} \right\rfloor = \hat{r}^-,$$

a contradiction.

$\square$

## Appendix B. Proofs for Section 6.2

The proof of Theorem B.1 proceeds by a fairly straightforward reduction from Plurality-CCAC. In contrast, the remaining proofs in this section employ reductions from X3C, and are quite technical.

**Theorem B.1.** *For each fixed $k$, $k \ge 1$, constructive control by adding candidates for $k$-Approval is NP-complete.*

*Proof.* For $k = 1$, $k$-Approval is simply Plurality and Plurality-CCAC has been shown to be NP-complete (Bartholdi et al., 1992). Thus, let us assume that $k \geq 2$.

Clearly, $k$-Approval-CCAC is in NP. To prove hardness, we give a reduction from Plurality-CCAC to $k$-Approval-CCAC. Given an instance $I = (C, A, \mathcal{R}, p, t)$ of Plurality-CCAC with $\mathcal{R} = (R_1, \ldots, R_n)$ (see Definition 6.1), we build an instance $\hat{I} = (\hat{C}, A, \hat{\mathcal{R}}, p, t)$ of $k$-Approval-CCAC as follows:

1. If $|\mathcal{R}| = 1$, we solve $I$ in polynomial time (which is easy to do in this case) and output a fixed instance of $k$-Approval-CCAC with the same answer.

2. If $n = |\mathcal{R}| > 1$, we set $D = \{d_{i,j} \mid 1 \leq i \leq n, 1 \leq j \leq k-1\}$ and let $\hat{C} = C \cup D$. For $i = 1, \ldots, n$, in the $i$-th preference order $\hat{R}_i$ the candidates $d_{i,1}, \ldots, d_{i,k-1}$ are ranked in positions $1, \ldots, k-1$, and the candidate that was ranked first in $R_i$ is ranked in position $k$. We set $\hat{\mathcal{R}} = (\hat{R}_1, \ldots, \hat{R}_n)$.

Clearly, if $n = 1$, the reduction works correctly. Now suppose $n > 1$. For each $A' \subseteq A$, and each candidate $c \in C \cup A'$, the Plurality score of $c$ in $(C \cup A', \mathcal{R})$ is the same as his $k$-Approval score in $(\hat{C} \cup A', \hat{\mathcal{R}})$, while the $k$-Approval score of any $d \in D$ in $(\hat{C} \cup A', \hat{\mathcal{R}})$ is 1. To complete the proof, it remains to observe that if some candidate $c^*$ is a unique winner of a Plurality election with at least two voters, then $c^*$ receives at least two points. $\qquad \square$

**Theorem B.2.** *Constructive control by adding candidates for Plurality with Runoff is NP-complete.*

*Proof.* The problem is clearly in NP. Our hardness reduction is from X3C. Let $I = (G, \mathcal{S})$ be an input instance of X3C, where $G = \{g_1, \ldots, g_{3K}\}$ is the ground set and $\mathcal{S} = \{S_1, \ldots, S_M\}$ is a family of 3-element subsets of $G$.

We construct an instance of CCAC for Plurality with Runoff as follows. We let the candidate set be $C = \{p, u, w\}$, and we let the set of spoiler candidates be $A = \{a_1, \ldots, a_M\}$. The preference profile $\mathcal{R}$ consists of $6K + 20$ preference orders $R_1, \ldots, R_{6K+20}$, described as follows. For $i = 1, \ldots, 3K$, let $A_i = \{a_j \in A \mid g_i \in S_j\}$. Preference orders $R_i$ and $R_{3K+i}$, $i = 1, \ldots, 3K$, are given by

$$A_i \succ u \succ A \setminus A_i \succ p \succ w \quad \text{and} \quad A_i \succ w \succ A \setminus A_i \succ p \succ u,$$

respectively. There are also 7 voters whose preference order is $p \succ A \succ u \succ w$, 7 voters whose preference order is $u \succ A \succ p \succ w$, and 6 voters whose preference order is $w \succ A \succ p \succ u$. Finally, we set $t = K$.

We claim that $p$ can become the unique winner of election $(C, \mathcal{R})$ by adding at most $t = K$ spoiler candidates from $A$ if and only if $I$ is a "yes"-instance of X3C.

Let us first consider Plurality scores of the candidates in $(C, \mathcal{R})$. We have $Sc_P(u) = 3K + 7$, $Sc_P(w) = 3K + 6$, and $Sc_P(p) = 7$. The runoff is between $u$ and $w$, and thus $p$ does not win.

Now, consider some subset $A'$ of $A$ and an election $(C \cup A', \mathcal{R})$. Let $Sc'_P(c)$ denote the Plurality score of a candidate $c \in C \cup A'$ in election $(C \cup A', \mathcal{R})$. We have $Sc'_P(p) = 7$, $Sc'_P(u) \geq 7$, $Sc'_P(w) = Sc'_P(u) - 1$, and, moreover, $Sc'_P(a_i) \leq 6$ for any $a_i \in A'$. This implies that no candidate from $A'$ can ever participate in the runoff. Thus, depending on the participating spoiler candidates, the following runoff scenarios are possible:

1. $Sc'_P(u) \geq 9$. Then $Sc'_P(w) \geq 8$, and the runoff is between $u$ and $w$.

2. $Sc'_P(u) = 8$. Then $Sc'_P(w) = 7$, and the runoff is between $u$ and either $p$ or $w$.

3. $Sc'_P(u) = 7$. Then $Sc'_P(w) = 6$, and the runoff is between $u$ and $p$.

As we use the parallel-universe tie-breaking rule, and more than half of the voters prefer $p$ to $u$, $p$ is the unique winner of the election if and only if $Sc'_P(u) = 7$. That is, $p$ can be made the unique winner of the election by adding at most $t$ candidates if and only if it is possible to choose a subset $A'$ of $A$ such that $|A'| \leq K$ and $u$ receives no Plurality points from the first $6K$ voters in election $(C \cup A', \mathcal{R})$. Clearly, such a set $A'$ has the property that in every preference order among $R_1, \ldots, R_{3K}$ some member of $A'$ is ranked above $u$. As $|A'| \leq K$, this is possible if and only if the collection $\mathcal{S}' = \{S_i \mid a_i \in A'\}$ is an exact 3-cover of $G$. □

To prove that CCAC control for Borda is NP-complete as well, we need a tool to construct Borda votes conveniently.

**Lemma B.3.** *Let $C = \{c_1, \ldots, c_{2t-1}, d\}$, $t \geq 2$, be a set of candidates and let $A = \{a_1, \ldots, a_s\}$ be a set of spoiler candidates. Let $\ell_1, \ldots, \ell_{2t-1}$ be a sequence of nonnegative integers, and set $L = \sum_{i=1}^{2t-1} \ell_i$. Then there is a preference profile $\mathcal{R} = (R_1, \ldots, R_{2L})$ over $C \cup A$ such that for each $A' \subseteq A$ the Borda scores in the election $(C \cup A', \mathcal{R})$ are as follows:*

1. *For each $c_i \in C$, $Sc_B(c_i) = L(2|A'| + |C| - 1) + \ell_i$.*

2. *$Sc_B(d) = L(2|A'| + |C| - 1) - L$.*

3. *For each $a_i \in A'$, $Sc_B(a_i) \leq L(2|A'| + |C| - 1) - 2L$.*

*Moreover, the preference profile $\mathcal{R}$ is computable in time polynomial in $|C| + |A| + L$.*

*Proof.* For each $i = 1, \ldots, 2t - 1$, set $e = c_i$, renumber the candidates in $C \setminus \{d, e\}$ as $b_1, \ldots, b_{2t-2}$, and consider preference orders $R_{2i-1}$ and $R_{2i}$ given by

$$R_{2i-1} \;:\; b_1 \succ b_2 \succ \ldots \succ b_{t-1} \succ e \succ d \succ b_t \succ \ldots \succ b_{2t-2} \succ A,$$
$$R_{2i} \;:\; b_{2t-2} \succ b_{2t-3} \succ \ldots \succ b_t \succ e \succ d \succ b_{t-1} \succ \ldots \succ b_1 \succ A.$$

Let $\mathcal{R}^i = (R_{2i-1}, R_{2i})$. For any given $A' \subseteq A$, in election $(C \cup A', \mathcal{R}^i)$ each $b_i \in C \setminus \{d, e\}$ receives $2|A'| + |C| - 1$ points, $e$ receives $2|A'| + |C|$ points, and $d$ receives $2|A'| + |C| - 2$ points. Moreover, each candidate $a_i$ in $A'$ receives at most $2|A'| - 2$ points.

Our preference profile $\mathcal{R}$ has $\ell_i$ copies of $R_{2i-1}$ and $\ell_i$ copies of $R_{2i}$ for each $i = 1, \ldots, 2t - 1$. It is easy to see that it satisfies the condition of the lemma. □

**Theorem B.4.** *Constructive control by adding candidates for Borda is NP-complete.*

*Proof.* It is easy to see that the CCAC control problem for Borda is in NP. We will now show that this problem is NP-hard by giving a reduction from X3C.

Let $(G, \mathcal{S})$ be our input instance of X3C, where $G = \{g_1, \ldots, g_{3K}\}$ is the ground set and $\mathcal{S} = \{S_1, \ldots, S_M\}$ is a collection of 3-element subsets of $G$. We assume without loss of generality that $K$ is even and $K > 2$; this can be achieved, e.g., by duplicating the instance.

We construct an instance of our problem as follows. The set of registered candidates $C$ is $\{p, d\} \cup G$ (we will sometimes refer to $p$ as $g_{3K+1}$) and the set of spoiler candidates is $A = \{a_1, \ldots, a_M\}$. The preference profile $\mathcal{R}$ consists of two parts, $\mathcal{R}'$ and $\mathcal{R}''$. We first describe $\mathcal{R}'$, and define $\mathcal{R}''$ based on the number of Borda points the candidates get from $\mathcal{R}'$. For each $i = 1, \ldots, M$, the preference profile $\mathcal{R}'$ contains exactly one preference order

$$A \setminus \{a_i\} \succ S_i \succ p \succ a_i \succ G \setminus S_i \succ d.$$

For each candidate $c \in C$, let $Sc_B'(c)$ be the number of Borda points that $c$ gets from $\mathcal{R}'$, assuming that no candidate from $A$ participates in the election. Set $S = \sum_{c \in C} Sc_B'(c)$. Observe that irrespective of what subset $A'$ of spoiler candidates participates in the election, each spoiler candidate $a_i$ gets at most $T = M(3K + M + 1)$ points from $\mathcal{R}'$. Also, in an election with the set of alternatives $C \cup A'$, candidate $p$ gets $Sc_B'(p) + |A'|$ points from $\mathcal{R}'$, while each $g_i \in G$ gets $Sc_B'(g_i) + |\{a_j \mid a_j \in A' \wedge g_i \in S_j\}|$ points from $\mathcal{R}'$.

For each $i = 1, \ldots, 3K$, we set

$$\ell_i = T + S - Sc_B'(g_i) - 2,$$

and for $p = g_{3K+1}$ we set

$$\ell_{3K+1} = T + S - Sc_B'(p) - K.$$

We obtain $\mathcal{R}''$ by applying Lemma B.3 to the candidate set $C$ (where each $g_i$ takes the role of $c_i$ and $p$ takes the role of $c_{3K+1}$), the spoiler candidate set $A$, and the sequence $\ell_1, \ldots, \ell_{3K+1}$; note that here we use the assumption that $K$ is even (and hence $|C|$ is even). Finally, we set $t = K$.

Let $L = \sum_{i=1}^{3K+1} \ell_i$, and for any $A' \subseteq A$ set $f(A') = L(2|A'| + |C| - 1) + T + S$. For any $A' \subseteq A$, in the election $(C \cup A', \mathcal{R}' + \mathcal{R}'')$ the candidates have the following Borda scores:

1. $Sc_B(p) = f(A') - K + |A'|$.

2. For each $g_i \in G$, we have $Sc_B(g_i) = f(A') - 2 + |\{a_j \mid a_j \in A' \wedge g_i \in S_j\}|$.

3. Irrespective of the choice of $A'$, for each $a_i \in A'$ we have $Sc_B(a_i) < Sc_B(p)$ and $Sc_B(d) < Sc_B(p)$.

Now it is easy to see that $p$ is not a Borda winner in the election $(C, \mathcal{R}' + \mathcal{R}'')$. Let us assume that there is a subset $A'$ of the spoiler candidates such that $|A'| \leq K$ and $p$ is the unique winner of election $(C \cup A', \mathcal{R}' + \mathcal{R}'')$. By the argument above, $A'$ is non-empty. Now, for any $a_j \in A'$ and any $g_i \in S_j$ we have $Sc_B(g_i) \geq f(A') - 1$. Therefore, for $p$ to be the unique winner of $(C \cup A', \mathcal{R}' + \mathcal{R}'')$, it has to be the case that $Sc_B(p) \geq f(A')$, i.e., $|A'| = K$, and, furthermore, for each $g_i \in G$ there is at most one $a_j \in A'$ such that $g_i \in S_j$. Hence, the collection $\mathcal{S}' = \{S_j \mid a_j \in A'\}$ consists of $K$ non-overlapping sets of size 3, i.e., it is an exact cover of $G$. Conversely, it is easy to see that if $\mathcal{S}'$ is an exact cover of $G$ by sets from $\mathcal{S}$ and $A' = \{a_j \mid S_j \in \mathcal{S}'\}$, then $p$ is the unique winner of $(C \cup A', \mathcal{R}' + \mathcal{R}'')$. This completes the proof. □

We remark that certain aspects of control under Borda have already been studied by Russel (2007); in fact, Russel mentions the idea of cloning in his work, but does not provide any results for cloning or CCAC control under Borda.

**Theorem B.5.** *Constructive control by adding candidates for Veto is NP-complete.*

*Proof.* The problem is clearly in NP. To show NP-hardness, we give a reduction from X3C. Let $(G, \mathcal{S})$ be an input instance of X3C, where $G = \{g_1, \ldots, g_{3K}\}$ is the ground set and $\mathcal{S} = \{S_1, \ldots, S_M\}$ is a family of 3-element subsets of $G$. Without loss of generality we can assume that $K > 2$. For each $i = 1, \ldots, 3K$, let $\ell_i = |\{S_j \in \mathcal{S} \mid g_i \in S_j\}|$.

We form the following instance of our problem. Our set of registered candidates is $C = G \cup \{p\}$, and the set of spoiler candidates is $A = \{a_1, \ldots, a_M\}$. The preference profile $\mathcal{R}$ consists of four subprofiles $\mathcal{R}^1$, $\mathcal{R}^2$, $\mathcal{R}^3$, and $\mathcal{R}^4$.

$\mathcal{R}^1$: $\mathcal{R}^1$ contains $4M$ preference orders in $M$ groups of four, one group for each set in $\mathcal{S}$. If $S_i = \{g_{i_1}, g_{i_2}, g_{i_3}\}$, the $i$-th group contains the following four preference orders:

$$A \setminus \{a_i\} \succ p \succ G \setminus \{g_{i_1}\} \succ g_{i_1} \succ a_i,$$
$$A \setminus \{a_i\} \succ p \succ G \setminus \{g_{i_2}\} \succ g_{i_2} \succ a_i,$$
$$A \setminus \{a_i\} \succ p \succ G \setminus \{g_{i_3}\} \succ g_{i_3} \succ a_i,$$
$$A \setminus \{a_i\} \succ G \succ p \succ a_i.$$

$\mathcal{R}^2$: $\mathcal{R}^2$ contains $3K$ groups of voters, where the $i$-th group consists of $M - \ell_i$ preference orders, i.e., $|\mathcal{R}^2| = \sum_{i=1}^{3K}(M - \ell_i) = 3KM - \sum_{i=1}^{3K}\ell_i = 3M(K-1)$. For each $i = 1, \ldots, 3K$, each of the preference orders in the $i$-th group is given by $A \succ p \succ G \setminus \{g_i\} \succ g_i$.

$\mathcal{R}^3$: $\mathcal{R}^3$ contains $K - 2$ preference orders of the form $A \succ G \succ p$.

$\mathcal{R}^4$: $\mathcal{R}^4$ contains $(3K + 1)M^2$ preference orders. For each $i = 1, \ldots, M$, $j = 1, \ldots, 3K$, there are $M$ preference orders of the form $p \succ G \setminus \{g_j\} \succ g_j \succ A \setminus \{a_i\} \succ a_i$ and for each $i = 1, \ldots, M$, there are $M$ preference orders of the form $G \succ p \succ A \setminus \{a_i\} \succ a_i$.

Intuitively, $\mathcal{R}^1$ models the input X3C instance, $\mathcal{R}^2$ ensures that within $\mathcal{R}^1$ and $\mathcal{R}^2$ all candidates receive the same number of points (assuming only candidates from $C$ participate), $\mathcal{R}^3$ models the constraint that the added candidates must correspond to a cover, and $\mathcal{R}^4$ ensures that none of the spoiler candidates can become a winner. Let us make these observations more formal by calculating the scores of candidates, assuming that the set of candidates is $C \cup A'$ for some $A' \subseteq A$.

For each $j = 1, \ldots, 3K$, let $t(A', g_j)$ be the number of sets $S_i \in \mathcal{S}$ such that $g_j \in S_i$ and $a_i \notin A'$. Candidate $g_j$ receives $|\mathcal{R}^1| - \ell_j + t(A', g_j)$ points from $\mathcal{R}^1$, $|\mathcal{R}^2| - (M - \ell_j)$ points from $\mathcal{R}^2$, and $|\mathcal{R}^3|$ points from $\mathcal{R}^3$. Thus, in total $g_j$ receives

$$4M - \ell_j + t(A', g_j) + 3M(K-1) - (M - \ell_j) + (K-2) = 3KM + t(A', g_j) + (K-2)$$

points from voters in $\mathcal{R}_1 + \mathcal{R}_2 + \mathcal{R}_3$. Similarly, $p$ receives $3M + |A'| + |\mathcal{R}^2| = 3KM + |A'|$ points from $\mathcal{R}^1 + \mathcal{R}^2 + \mathcal{R}^3$ and each $a_i \in A'$ receives $|\mathcal{R}^1| - 4 + |\mathcal{R}^2| + |\mathcal{R}^3| = 4M - 4 + 3M(K-1) + (K-2) = M(3K+1) + (K-6)$ points from voters in $\mathcal{R}^1 + \mathcal{R}^2 + \mathcal{R}^3$.

It remains to calculate how many points each candidate receives from $\mathcal{R}^4$. Note that there are $(3K+1)M^2$ preference orders in $\mathcal{R}^4$ and each $a_i \in A'$ is ranked last in at least $(3K+1)M$ of them. Thus, each $a_i \in A'$ receives at most $(3K+1)M^2 - (3K+1)M$ points from $\mathcal{R}^4$. On the other hand, if $A' \neq \emptyset$, each candidate in $C$ receives exactly $(3K+1)M^2$ points from $\mathcal{R}^4$, whereas if $A' = \emptyset$, each candidate in $C$ receives $3KM^2$ points from $\mathcal{R}^4$.

We set $t = K$, and ask whether $p$ can be made the unique winner by adding at most $t$ spoiler candidates. It is easy to see that if $A' = \emptyset$, candidate $p$ loses to all candidates in $G$, and thus is not the unique winner of the election. Now suppose that $A' \neq \emptyset$. Set $F = (3K+1)M^2 + 3KM + K - 1$. The candidates have the following scores:

1. $Sc_V(p) = F + |A'| - K + 1$,

2. for each $g_j \in G$, we have $Sc_V(g_j) = F - 1 + t(A', g_j)$, and

3. for each $a_i \in A'$, we have $Sc_V(a_i) \leq F - 3KM - 5$.

Thus, clearly no member of $A'$ is a winner. Let us now assume that $0 < |A'| \leq K$ and $p$ is the unique winner of the election. Since $t(A', g_j) \geq 1$ for at least one $g_j \in G$, it holds that some $g_j \in G$ has at least $F$ points, and therefore the score of $p$ is at least $F + 1$. This implies that $|A'| = K$. Further, for each $g_j \in G$ it must be the case that $t(A', g_j) \leq 1$. Since $|A'| = K$, this means that the collection $\{S_i \mid a_i \in A'\}$ is an exact cover of $G$ by sets from $\mathcal{S}$.

It is easy to see that the converse direction is also true. Thus, it is possible to ensure that $p$ is a winner of our election by adding at most $k$ candidates from $A'$ if and only if $(G, \mathcal{S})$ is a "yes"-instance of X3C. $\qquad \square$

**Theorem 6.2.** *There is a voting rule for which constructive control by adding candidates is in P, but UC $0^+$-Cloning is NP-hard.*

*Proof.* The idea of the proof, common to several results of this type (see, e.g., the paper of Faliszewski et al., 2009, where the authors construct a voting rule for which the problem of bribery is in P, but the problem of manipulation is NP-hard), is to embed an NP-complete problem into the winner determination procedure of the newly constructed voting rule.

Let $L$ be an NP-complete language over the alphabet $\Sigma = \{0, 1\}$. We will soon provide some further assumptions regarding $L$ (all easily satisfied), but before doing so, we describe how an election can encode a pair of strings over $\Sigma$.

Fix an election $E = (A, \mathcal{R})$ with $\mathcal{R} = (R_1, \ldots, R_\ell, R_{\ell+1})$. We can assume without loss of generality that $A = \{c_1, \ldots, c_m\}$ and that the last voter ranks the candidates as $c_1 \succ_{\ell+1} c_2 \succ_{\ell+1} \cdots \succ_{\ell+1} c_m$. We say that $E$ encodes two length-$\ell$ binary strings, $x = x_1 \ldots x_\ell$ and $y = y_1 \ldots y_\ell$, if the following conditions hold:

1. $\ell > 0$ and $m \geq 4$.

2. Each voter ranks $c_1$ and $c_2$ in the top two positions.

3. For each $i = 1, \ldots, \ell$, if $y_i = 0$ then $c_1 \succ_i c_2$ and if $y_i = 1$ then $c_2 \succ_i c_1$.

4. For each $i = 1, \ldots, \ell$, if $x_i = 0$ then $c_3 \succ_i c_4 \succ_i \cdots \succ_i c_m$ and if $x_i = 1$ then $c_m \succ_i c_{m-1} \succ_i \cdots \succ_i c_3$.

Otherwise $E$ does not encode any strings. Note that the last preference order $R_{\ell+1}$ is special, as it defines the roles of the candidates; also, by requiring $\ell > 0$ we explicitly forbid encoding a pair of two empty strings.

By definition of NP and by basic properties of NP-complete languages, we can assume that language $L$ admits a polynomial-time algorithm $\mathcal{B}$ such that for every binary string $x$ of length $\ell$ it holds that $x \in L$ if and only if there is a binary string $y$ of length $\ell$ such that $\mathcal{B}(x, y)$ accepts.[6] Further, we assume that $L$ does not contain the empty string and that it does not contain any all-0 strings.

We will now define our voting rule, which we call $\mathcal{L}$. Let $E = (A, \mathcal{R})$ be an election. If $E$ encodes strings $x$ and $y$ such that $\mathcal{B}(x, y)$ accepts (that is, such that $x \in L$ and $y$ witnesses that this is the case), then all candidates in $A$ are winners. Otherwise, the winner is the candidate that is ranked last by the last voter.

First, we claim that $\mathcal{L}$-CCAC is in P. Let $I = (E, p, t)$ be an instance of $\mathcal{L}$-CCAC, where $E = (C \cup A, \mathcal{R})$, $C = \{c_1, \ldots, c_m\}$, $A = \{a_1, \ldots, a_{m'}\}$, $\mathcal{R} = (R_1, \ldots, R_n)$, $p \in C$ is the preferred candidate, and $t \in \mathbb{Z}^+$ is the bound on the number of candidates that we can add. If $p$ is a winner prior to adding candidates, we accept. Note that if $p$ is not ranked last by the last voter, we cannot change that by adding candidates from $A$. Therefore, to make $p$ a winner, we have to add candidates so as to influence the strings $x, y$ encoded by the election or to move the election from the state in which it does not encode any strings to the state where it does. However, if this goal can be achieved by adding some $s \leq t$ candidates, then it can also be achieved by adding at most 4 candidates. Indeed, suppose that we add some candidates $a_{i_1}, \ldots, a_{i_s}$, $4 < s \leq t$, so that $a_{i_1} \succ_n a_{i_2} \succ_n \cdots \succ_n a_{i_s}$ and the resulting election encodes some strings $x$ and $y$. It is easy to see that if we now remove the candidates $a_{i_5}, \ldots, a_{i_s}$, the resulting election also encodes the same strings $x$ and $y$. Thus, to test if it is possible to make $p$ a winner by adding candidates from $A$, it suffices to try adding all subsets of $A$ of size at most 4, and, for each of them, verify whether the resulting election encodes two strings $x$ and $y$ such that $\mathcal{B}(x, y)$ accepts. Clearly, this can be done in polynomial time.

On the other hand, UC $0^+$-Cloning is NP-complete for $\mathcal{L}$ (in fact, this remains true for ZC $0^+$-Cloning or any other cost model that allows adding at least one clone). We give a reduction from $L$. Let $x$ be our input binary string of length $\ell$; we can assume without loss of generality that $\ell > 0$. We create an election $E = (A', \mathcal{R})$ where $A' = \{a, c_1, c_2, c_3\}$ is the set of candidates and $\mathcal{R} = (R_1, \ldots, R_\ell, R_{\ell+1})$ is the preference profile, where

1. $R_{\ell+1}$ is given by $a \succ_{\ell+1} c_1 \succ_{\ell+1} c_2 \succ_{\ell+1} c_3$.

2. For each $i = 1, \ldots, \ell$, if $x_i = 0$ then $R_i$ is given by $a \succ_i c_1 \succ_i c_2 \succ_i c_3$ and if $x_i = 1$ then $R_i$ is given by $a \succ_i c_3 \succ_i c_2 \succ_i c_1$.

It is easy to see that either this election encodes a pair of strings $(0^\ell, 0^\ell)$ or it does not encode any strings. We pick $a$ as our preferred candidate; note that $a$ is not ranked last by the last voter.

Suppose that $x \in L$, i.e., there exists a string $y$, $|y| = \ell$, such that $\mathcal{B}(x, y)$ accepts. Then we can clone $a$ into $a^{(1)}$ and $a^{(2)}$ and ask the voters to order the clones so that the top two

---

positions in their preference orders encode $y$. Clearly, in the resulting election all candidates win.

Conversely, suppose that it is possible to clone some candidates in $A'$ so as to make $a$ a winner (i.e., so that all candidates are winners). Let $E'$ be the election after cloning. It must be the case that $E'$ encodes two strings, $x'$ and $y'$, such that $\mathcal{B}(x', y')$ accepts. For this to happen, each voter must rank clones of $a$ in the first two positions and the clones of $c_1$, $c_2$ and $c_3$ in the remaining positions. This implies that $x' = x$. Hence, if there is a $0^+$-successful cloning for $E$, then there is one that replaces $a$ with two clones, $a^{(1)}$ and $a^{(2)}$, and asks the voters to rank the clones so that they encode a string $y$ with the property that $\mathcal{B}(x, y)$ accepts.

Our reduction can be computed in polynomial time and thus the proof is complete. $\quad\square$